\documentclass[twocolumn,tighten,twocolappendix]{aastex63}

\usepackage{amsmath}
\usepackage[caption=false]{subfig}

\received{2020 April 16}
\revised{2020 June 15}
\accepted{2020 June 17}
\published{2020 August 7}
\submitjournal{\apj}

\shorttitle{Statistical detection of 21~cm Forest During Cosmic Reionization}
\shortauthors{Thyagarajan}
\graphicspath{{./}{figures/}}
\begin{document}

\title{Statistical Detection of IGM Structures during Cosmic Reionization using Absorption of the Redshifted 21~cm line by H~{\sc i} against Compact Background Radio Sources}

%% The \author command is the same as before except it now takes an optional
%% argument which is the 16 digit ORCID. The syntax is:
%% \author[xxxx-xxxx-xxxx-xxxx]{Author Name}
%%
%% This will hyperlink the author name to the author's ORCID page. Note that
%% during compilation, LaTeX will do some limited checking of the format of
%% the ID to make sure it is valid. If the "orcid-ID.png" image file is 
%% present or in the LaTeX pathway, the OrcID icon will appear next to
%% the authors name.
%%
%% Use \affiliation for affiliation information. The old \affil is now aliased
%% to \affiliation. AASTeX v6.3 will automatically index these in the header.
%% When a duplicate is found its index will be the same as its previous entry.
%%
%% Note that \altaffilmark and \altaffiltext have been removed and thus 
%% can not be used to document secondary affiliations. If they are used latex
%% will issue a specific error message and quit. Please use multiple 
%% \affiliation calls for to document more than one affiliation.
%%
%% The new \altaffiliation can be used to indicate some secondary information
%% such as fellowships. This command produces a non-numeric footnote that is
%% set away from the numeric \affiliation footnotes.  NOTE that if an
%% \altaffiliation command is used it must come BEFORE the \affiliation call,
%% right after the \author command, in order to place the footnotes in
%% the proper location.

% \correspondingauthor{Nithyanandan Thyagarajan}
\email{t\_nithyanandan@nrao.edu, nithyanandan.t@gmail.com}

\author[0000-0003-1602-7868]{Nithyanandan Thyagarajan}
\altaffiliation{Nithyanandan Thyagarajan is a Jansky fellow of the National Radio Astronomy Observatory.}
\affiliation{National Radio Astronomy Observatory, 1003 Lopezville Rd, Socorro, NM 87801, USA}

\begin{abstract}

Detecting neutral hydrogen structures in the intergalactic medium (IGM) during cosmic reionization via absorption (21~cm forest) against a background radiation is considered independent and complementary to the three-dimensional tomography and power spectrum techniques. The direct detection of this absorption requires very bright ($\gtrsim 10$--100~mJy) background sources at high redshifts ($z\gtrsim 8$), which are evidently rare; very long times of integration; or instruments of very high sensitivity. This motivates a statistical one-dimensional (1D) power spectrum approach along narrow sightlines but with fainter background objects ($\sim 1$--10~mJy), which are likely to be more abundant and significant contributors at high redshifts. The 1D power spectrum reduces cosmic variance and improves sensitivity especially on small spatial scales. Using standard radiative transfer and fiducial models for the instrument, the background sources, and the evolution of IGM structures during cosmic reionization, the potential of the 1D power spectrum along selected narrow directions is investigated against uncertainties from thermal noise and the chromatic synthesized point spread function (PSF) response. Minimum requirements on the number of high-redshift background sources, the telescope sensitivity, and the PSF quality are estimated for a range of instrumental, background source, and reionization model parameters. The 1D power spectrum is intrinsically stronger at higher redshifts. A $\sim 1000$~hr observing campaign targeting $\sim 100$ narrow sightlines to radio-faint, high-redshift background objects with modern radio telescopes, especially the Square Kilometre Array, can detect the 1D power spectrum on a range of spatial scales and redshifts, and potentially discriminate between models of cosmic reionization. 

\end{abstract}

%% Keywords should appear after the \end{abstract} command. 
%% See the online documentation for the full list of available subject
%% keywords and the rules for their use.
% \keywords{editorials, notices --- 
% miscellaneous --- catalogs --- surveys}

\section{Introduction} \label{sec:intro}

An important phase in the history of the universe is the cosmic-scale reionization process following the appearance of the first self-luminous objects in the universe such as the first stars and galaxies. Spanning multiple epochs that are known as {\it cosmic dawn} (CD), epoch of heating, and epoch of reionization (EoR), etc., this phase signifies an important period of nonlinear growth of cosmic structures and processes that shaped the astrophysical evolution of the universe. The detection of the most abundant element, namely neutral hydrogen (H~{\sc i}), not only promises to provide direct constraints on the structures and astrophysical processes during this epoch \citep{sun72,sco90,mad97,toz00,ili02} but also appears viable with the advancement of modern radio telescopes that target the 21~cm spectral line associated with the electron spin-flip transition in H~{\sc i}, which is redshifted to 50--200~MHz frequencies. 

One approach is to detect the H~{\sc i} structures in the {\it intergalactic medium} (IGM) on different scales, either via imaging or statistical techniques employing second-order (e.g. variance, power spectrum) and higher-order moments (e.g. bispectrum, skewness, kurtosis) using large interferometer arrays at low radio frequencies such as the Low Frequency Array \citep[LOFAR;][]{van13}, the Murchison Widefield Array \citep[MWA;][]{bow13,tin13,bea19}, the Precision Array for Probing the Epoch of Reionization \citep[PAPER;][]{par10}, the Long Wavelength Array \citep[LWA;][]{ell09}, the Hydrogen Epoch of Reionization Array \citep[HERA\footnote{\url{https://reionization.org}};][]{deb17,thy16,neb16,ewa16,patra18}, and the Square Kilometre Array \citep[SKA\footnote{\url{https://www.skatelescope.org}};][]{braun19}. Current instruments such as LOFAR, MWA, and HERA will have sufficient sensitivity only for a statistical detection \citep{bea13,thy13} using either a three-dimensional (3D) power spectrum approach \citep{mor04,mor05,mcq06,lid08} or using the bispectrum phase that is more robust to the calibration challenges \citep{thy18,thy20a,thy20b,car18,car20}. The 3D tomography of the IGM structures at these high redshifts will require even powerful instruments such as the SKA. 

An alternate approach is to observe the redshifted 21~cm spectra in absorption against compact sources of radiation such as quasars and Active Galactic Nuclei (AGN), star-forming radio galaxies, or radio afterglows of Gamma Ray Bursts (GRB) in the background that shine through these cosmic epochs \citep{car02,car04,car07,fur06a,mac12,cia13,cia15a,cia15b,chapman+2019}. Such a detection in individual spectral channels of narrow width requires either extremely sensitive instruments (or very long observing times) or very bright background sources, typically $\gtrsim 10$--100~mJy \citep{car02,fur06a,mac12,cia13,cia15b}. Such bright radio AGNs are evidently rare at high redshifts \citep{sax17,bolgar18}, at least based on the current sample of known quasars which are predominantly optical selections \citep[see for example][]{banados16}. Statistical methods based on the change of variance in the absorbed regions of the spectrum have also been proposed \citep{car02,mac12}. The statistical effect of the presence of a distribution of background quasars on the aforementioned standard 3D power spectrum has been explored \citep{ewa14} and the modeling showed an excess power on small scales (large wavenumbers or $k$-modes). 

This paper is distinct in that it combines the advantage of targeting specific sightlines which are known to contain compact background objects (including those faint in radio) that dominate the background radiation relative to the CMB as well as obtains the additional sensitivity of a statistical approach, namely, the one-dimensional (1D) power spectrum of the spectra along the specific sightlines selected. Models indicate that at high redshifts, radio-faint ($\lesssim 10$~mJy) background sources such as AGNs are expected to be much more abundant \citep{hai04,sax17} than brighter objects, thus making a statistical approach valuable. In addition, relative to a direct detection of absorption features against a few rare background objects, the 1D power spectrum approach uses multiple sightlines and will significantly reduce cosmic variance, as well as provide sensitivity over potentially a wider range of spatial scales that may not be accessible to the former. Though related, this approach is distinct from that in \citet{ewa14} because the latter primarily explores the net effect of the redshifted 21~cm forest on the standard 3D power spectrum obtained in which most of the background radiation is still the CMB over the entire field of view. In such a scenario, the H~{\sc i} signal on the sky is present in both emission and absorption and will result in a lower average signal level in interferometric measurements, especially on the larger spatial scales. The 1D power spectrum approach here can avoid some of the challenges in the analysis of wide-field measurements that are typically used in the tomography and 3D power spectrum approaches. In addition to exploring the number count of high-redshift background radiation sources and the sensitivity requirements from modern low-frequency radio telescopes for signal detection, this paper also investigates the effect of sidelobes from the synthesized point spread function (PSF) and places constraints on the quality of the synthesized aperture, which is seldom addressed in redshifted 21~cm absorption studies.

The paper is organized as follows. \S\ref{sec:models} presents two models using simulations of the evolution of the H~{\sc i} structure that bracket a broad range of astrophysical parameters, models for the instrument and observations generically applicable to a variety of radio interferometer arrays, and a simple continuum model for the spectrum of compact background radiation sources. In \S\ref{sec:absorption}, a contextual review and a demonstration of the radiative transfer equations are presented using nominal instrument and observing parameters. \S\ref{sec:methods} focuses on the 1D power spectrum methodology but also briefly reviews a few methods for detecting the absorption features including the direct detection approach. In \S\ref{sec:N_gamma_required}, the requirement on the number of background sources is derived for a successful detection of the 1D power spectrum on all accessible spatial scales using nominal instrument and observing parameters. \S\ref{sec:array-sensitivity-required} explores the requirements on the array sensitivity and places the prospects of detection by various modern radio instruments in context. In \S\ref{sec:sidelobes}, the effect of the chromatic structure of the sidelobes from aperture synthesis on the 1D power spectrum is investigated, and specifications on the quality of the synthesized point spread function (PSF) are presented. The summary and conclusions are presented in \S\ref{sec:summary}. A brief review of the radio $K$-correction required between two cosmological frames is presented in Appendix \ref{sec:K-correction}. Appendix \ref{sec:optical-depth-PS} outlines a simple framework to estimate the 1D power spectrum of optical depth.

\section{Modeling} \label{sec:models}

In this paper, the following notation for the different coordinates in different reference frames will be adopted. The unit vector pointing along each sightline will be denoted by $\hat{\boldsymbol{s}}$, frequency by $\nu$, and redshift by $z$. Quantities such as brightness temperature, specific intensity, and flux density will depend on $\hat{\boldsymbol{s}}$, $\nu$ as well as on the reference frame that depends on $z$. For example, a background source of radiation will be described by its brightness temperature $T_\gamma(\hat{\boldsymbol{s}},\nu,z)$ in the reference frame at redshift $z$ but differently as $T_\gamma(\hat{\boldsymbol{s}},\nu,z=0)$ in the observer's frame at $z=0$. The CMB monopole temperature has no dependence on direction or frequency and will thus be simply denoted by $T_\textrm{CMB}(z) = (1+z)\,T_\textrm{CMB}(z=0)$, where $T_\textrm{CMB}(z=0)=2.725$~K. Quantities such as optical depth and spin temperature, $T_\textrm{s}\equiv T_\textrm{s}(\hat{\boldsymbol{s}},z)$ are primarily defined with respect to the characteristic 21~cm spectral line transition arising from the spin-flip of the electron in H~{\sc i} with rest frequency, $\nu_\textrm{r}$, at the appropriate redshift. Thus, their redshift and frequency are tied to each other as $\tau(\hat{\boldsymbol{s}},\nu,z)\equiv \tau_\nu(\hat{\boldsymbol{s}},\nu)\equiv \tau_z(\hat{\boldsymbol{s}},z=\frac{\nu_\textrm{r}}{\nu}-1)$.
Angular brackets around any quantity refer to the average of that quantity marginalized over all available $\hat{\boldsymbol{s}}$, $\langle\cdots\rangle \equiv \langle\cdots\rangle_{\hat{\boldsymbol{s}}}$. 

The modeling in this paper consists of three components, namely, the evolution of H~{\sc i} in the IGM during the pre-CD through post-EoR, a simple radio continuum model for the source of compact background radiation which could generically encompass any compact radio-emitting object at high $z$ including AGNs, star-forming galaxies and radio afterglows from GRBs, and an instrument model. Each of these components is described below. 

\subsection{Evolution of H~{\sc i} Structures in the IGM} \label{sec:21cmfast}

The redshift evolution of H~{\sc i} in the IGM at these cosmic epochs is provided by the 21cmFAST simulations \citep{mes11}. Specifically, two fiducial models named the {\tt BRIGHT GALAXIES} and the {\tt FAINT GALAXIES} models \citep{mes16}, which are publicly available as the Evolution of 21~cm Structure (EoS) datasets\footnote{\url{http://homepage.sns.it/mesinger/EOS.html}}, are used in this paper. These models represent two extreme choices for the minimum threshold of the virial temperature of the halo hosting the star-forming galaxies while matching the current constraints on the reionization and the cosmic star formation histories. This parameter significantly influences both the timing of the epochs and the typical bias of the dominant galaxies, and is thus representative of the largest variation in the 21~cm signal \citep{mes16,gre17b}. The {\tt BRIGHT GALAXIES} and the {\tt FAINT GALAXIES} models will be interchangeably referred to as the EoS models 1 and 2, respectively, hereafter. 

The simulated light-cone cubes for these models span 1.6~comoving Gpc (cGpc) in the transverse plane of the sky, and redshifts extend over the range $5\lesssim z \lesssim 80$ with uniform spacing in comoving distance along the line of sight. Each voxel is a cube of dimension 1.5625 comoving Mpc (cMpc) on each side. The simulations were obtained using the following cosmological parameters following the standard terminology: $H_0=67.8\,\textrm{km}\,\textrm{s}^{-1}\,\textrm{Mpc}^{-1}$ ($h=0.678$), $\Omega_m=0.308$, $\Omega_\Lambda=1-\Omega_\textrm{m}$, $\Omega_b\,h^2=0.02226$, $Y_\textrm{He}=0.245$, $w=-1$, and $\sigma_8=0.815$. In order to be compatible with $h=1$, the voxel dimensions were subsequently adjusted to $\delta r_x = \delta r_y = \delta r_z = 1.059375\,h^{-1}\,$cMpc. 

\begin{figure}
\includegraphics[width=\linewidth]{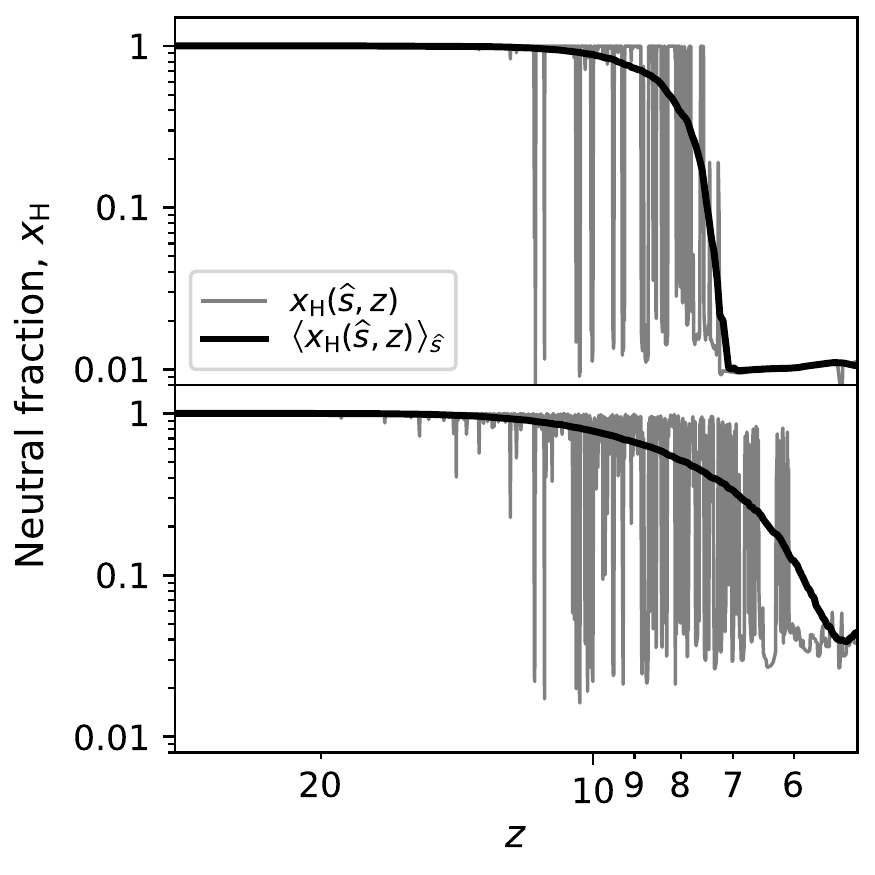}
\caption{The redshift evolution of neutral fraction obtained from 21cmFAST simulations along an arbitrary direction, $x_\textrm{H}(\hat{\boldsymbol{s}},z)$ (gray curve), and averaged over all directions, $\langle x_\textrm{H}(\hat{\boldsymbol{s}},z)\rangle$ (black curve) for the {\tt BRIGHT GALAXIES} model (EoS model 1) and the {\tt FAINT GALAXIES} model (EoS model 2) in the top and bottom panels, respectively.
\label{fig:neutral_fraction}}
\end{figure}

The simulations provide the redshift evolution of the neutral fraction of hydrogen, the fractional baryonic density fluctuations, and the 21~cm spin temperature of H~{\sc i} denoted by $x_\textrm{H}(\hat{\boldsymbol{s}}, z)$, $\delta_b(\hat{\boldsymbol{s}},z)$, and $T_\textrm{s}(\hat{\boldsymbol{s}},z)$, respectively. Figure~\ref{fig:neutral_fraction} shows the neutral fraction (sky-averaged in black and along an arbitrary sightline, $\hat{\boldsymbol{s}}$, in gray) as a function of redshift in the EoS models 1 (top) and 2 (bottom). The EoS model 1 exhibits a sharper depletion of the neutral fraction during the reionization process and a lower eventual neutral fraction relative to the EoS model 2.   

\begin{figure}
\includegraphics[width=\linewidth]{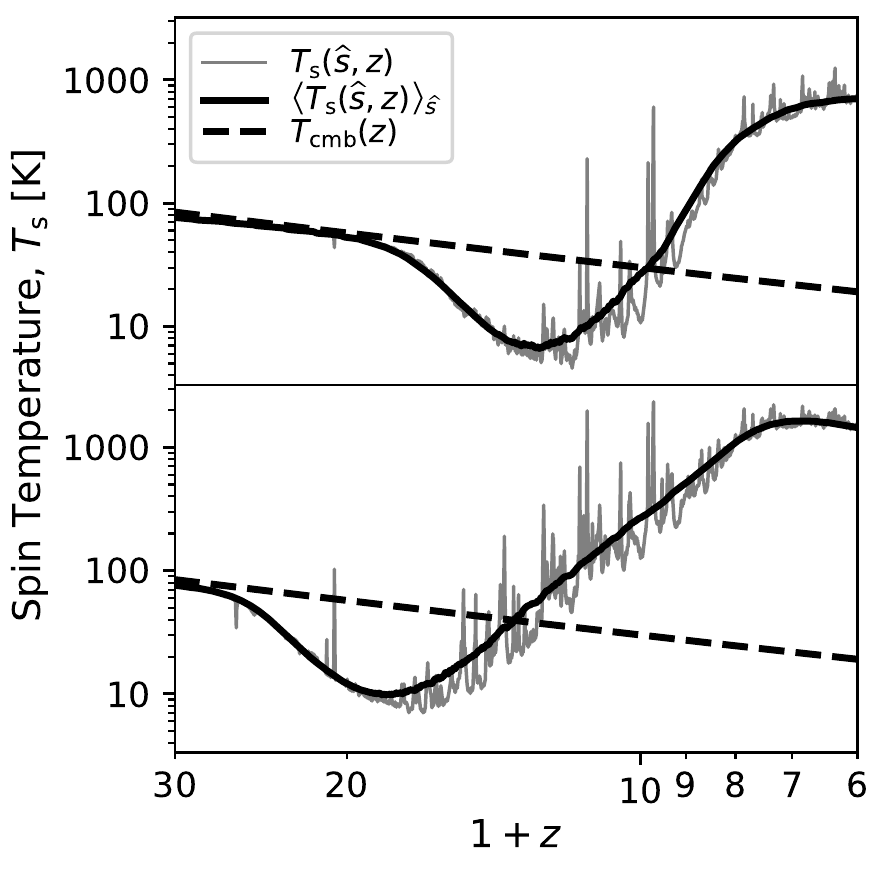}
\caption{The redshift evolution of the spin temperature, $T_\textrm{s}$, obtained from 21cmFAST simulations along an arbitrary direction, $T_\textrm{s}(\hat{\boldsymbol{s}},z)$ (solid gray curve), and averaged over all directions, $\langle T_\textrm{s}(\hat{\boldsymbol{s}},z)\rangle$ (solid black curve) for the {\tt BRIGHT GALAXIES} model (EoS model 1) and the {\tt FAINT GALAXIES} model (EoS model 2) in the top and bottom panels, respectively. The redshift evolution of $T_\textrm{cmb}$ (dashed black line) is shown for reference.
\label{fig:T_spin}}
\end{figure}

Figure~\ref{fig:T_spin} depicts the evolution of spin temperature, $T_\textrm{s}(\hat{\boldsymbol{s}},z)$, with redshift. The appearance of the first luminous objects ({\it cosmic dawn}), and the onset of the heating and reionization epochs happen relatively later in the EoS model 1. 

\begin{figure}
\includegraphics[width=\linewidth]{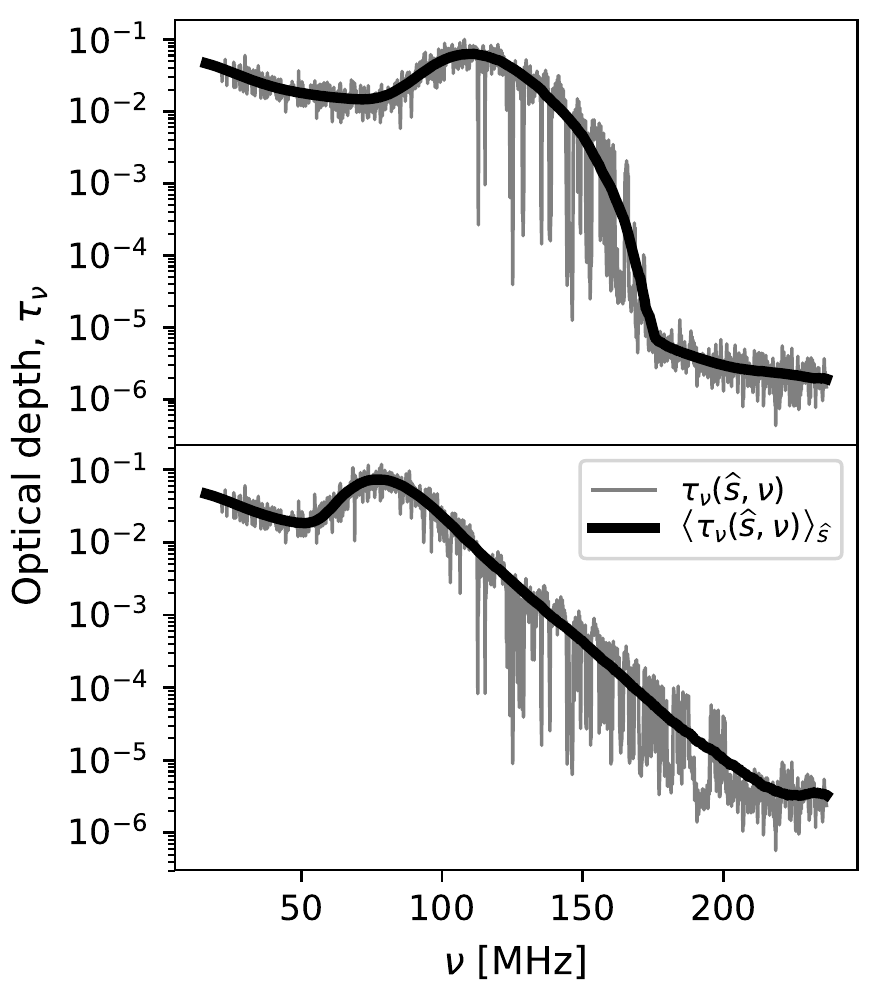}
\caption{The frequency (redshift) evolution of optical depth obtained from 21cmFAST simulations using Equation~(\ref{eqn:optical-depth}),  along an arbitrary direction, $\tau_\nu(\hat{\boldsymbol{s}},z)$ (gray), and averaged over all directions, $\langle \tau_\nu(\hat{\boldsymbol{s}},z)\rangle$ (black) for the {\tt BRIGHT GALAXIES} model (EoS model 1; top) and the {\tt FAINT GALAXIES} model (EoS model 2; bottom). Note that $\nu = \nu_\textrm{r}/(1+z)$.
\label{fig:optical_depth}}
\end{figure}

The optical depth to 21~cm radiation is estimated using \citep{field59a,san05}:
\begin{align}\label{eqn:optical-depth}
  \tau_z(\hat{\boldsymbol{s}}, z) &\equiv \tau_\nu\left(\hat{\boldsymbol{s}}, \nu=\nu_\textrm{r}/(1+z)\right) \nonumber\\
                                  &\approx 8.6\times 10^{-3}\left[1+\delta_b(\hat{\boldsymbol{s}}, z)\right]x_\textrm{H}(\hat{\boldsymbol{s}},z)\left[\frac{T_\textrm{CMB}(z)}{T_\textrm{s}(\hat{\boldsymbol{s}}, z)}\right]\nonumber\\
    &\quad \times \left(\frac{\Omega_b h^2}{0.02}\right)\left(\frac{1+z}{10}\,\frac{0.15}{\Omega_m h^2}\right)^{1/2}\left(\frac{h}{0.7}\right)^{-1}
\end{align}
Figure~\ref{fig:optical_depth} shows the optical depth for the EoS models in this study. The top and bottom panels represent EoS models 1 and 2, respectively. The gray line indicates the optical depth along an arbitrary sightline, $\hat{\boldsymbol{s}}$, while the black line denotes the optical depth evolution after averaging over the transverse plane of the sky, $\langle\tau_\nu(\hat{\boldsymbol{s}},\nu)\rangle$. From Equation~(\ref{eqn:optical-depth}), the optical depth is found to closely follow the evolution of the inverse of the spin temperature.

With a background radiation denoted by brightness temperature, $T_\gamma(\hat{\boldsymbol{s}},\nu,z)$, the radiative transfer equation \citep{Rybicki-Lightman} can be used to obtain the evolution of the net brightness temperature in the observed frame as \citep{pri12}:
\begin{align}\label{eqn:Tb_obs}
    T_\textrm{b}(\hat{\boldsymbol{s}},\nu) &= [T_\textrm{s}(\hat{\boldsymbol{s}},z)(1-e^{-\tau_\nu(\hat{\boldsymbol{s}},\nu)}) \nonumber\\
    &\qquad + T_\gamma(\hat{\boldsymbol{s}},\nu[1+z],z)e^{-\tau_\nu(\hat{\boldsymbol{s}},\nu)}] / (1+z) \\
    &\approx \frac{T_\textrm{s}(\hat{\boldsymbol{s}},z)-T_\gamma(\hat{\boldsymbol{s}},\nu[1+z],z)}{1+z}\tau_\nu(\hat{\boldsymbol{s}},\nu) \nonumber\\ 
    &\quad + T_\gamma(\hat{\boldsymbol{s}},\nu,z=0),
\end{align}
where the approximation in the last step resulted from assuming that $\tau_\nu(\hat{\boldsymbol{s}},\nu) \ll 1$, and from Equation~(\ref{eqn:temperature-redshift}) $T_\gamma(\hat{\boldsymbol{s}},\nu[1+z],z) = (1+z)\,T_\gamma(\hat{\boldsymbol{s}},\nu,z=0)$ (see \S\ref{sec:K-correction} for a generalized treatment of the radio K-correction for the specific brightness and temperature). Usually, it is assumed that the CMB provides the background radiation, $T_\gamma(\hat{\boldsymbol{s}},\nu,z) = T_\textrm{CMB}(z)$. Then, the observed brightness temperature contrast relative to $T_\textrm{CMB}$ is:
\begin{align}\label{eqn:Tb_obs_cmb}
    \delta T_\textrm{b}(\hat{\boldsymbol{s}},\nu) &\approx \frac{T_\textrm{s}(\hat{\boldsymbol{s}},z) -  T_\textrm{CMB}(z)}{1+z}\,\tau_\nu(\hat{\boldsymbol{s}},\nu),
\end{align}
where $z=\nu_\textrm{r}/\nu-1$. 

\begin{figure}
\includegraphics[width=\linewidth]{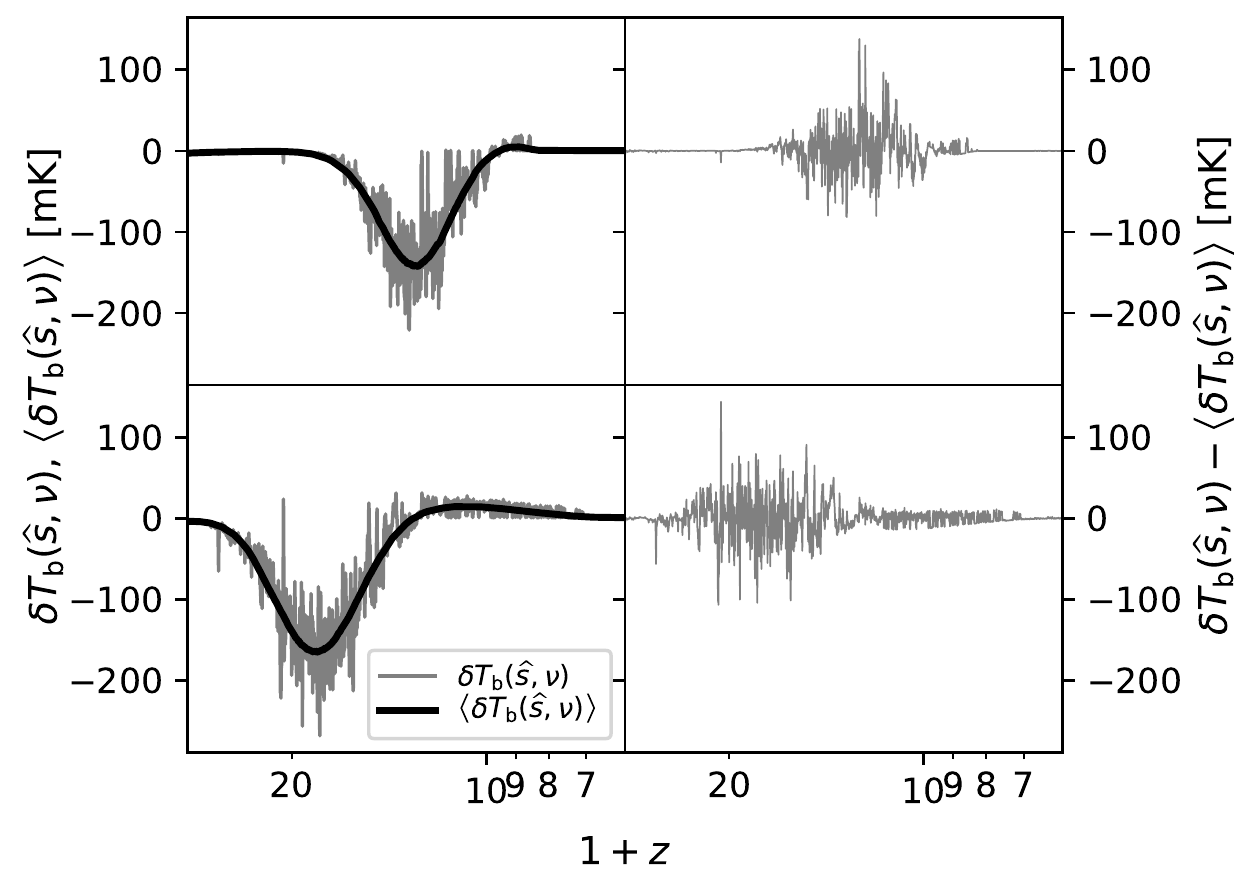}
\caption{{\it Top}: Redshift evolution of the brightness temperature contrast, $\delta T_\textrm{b}(\hat{\boldsymbol{s}},\nu)$ (along an arbitrary direction in gray) and $\langle \delta T_\textrm{b}(\hat{\boldsymbol{s}},\nu)\rangle$ (averaged over all directions in black) measured relative to $T_\textrm{CMB}(z=0)$ on the left subpanel for {\tt BRIGHT GALAXIES} model (EoS model 1). The right subpanel shows the brightness temperature fluctuations with the sky-averaged component (black) subtracted, $\delta T_\textrm{b}(\hat{\boldsymbol{s}},\nu) - \langle \delta T_\textrm{b}(\hat{\boldsymbol{s}},\nu)\rangle$ that will be observed by an interferometer. {\it Bottom}: Same as the top panel but for the {\tt FAINT GALAXIES} model (EoS model 2). Note that $\nu = \nu_\textrm{r}/(1+z)$.
\label{fig:predicted_dTb}}
\end{figure}

The gray and black curves in the left-hand panels of Figure~\ref{fig:predicted_dTb} show $\delta T_\textrm{b}(\hat{\boldsymbol{s}},\nu)$ for an arbitrary $\hat{\boldsymbol{s}}$ and the sky-averaged $\langle \delta T_\textrm{b}(\hat{\boldsymbol{s}},\nu)\rangle$ respectively, for the EoS models 1 (top) and 2 (bottom). Single antenna experiments such as EDGES \citep{bow10,bow18,monsalve17a,monsalve17b,monsalve18,monsalve19}, and SARAS \citep[SARAS;][]{patra13,patra15,sin17,sin18a,sin18b} aim to detect the sky-averaged (monopole) spectrum of the brightness temperature contrast, $\langle \delta T_\textrm{b}(\hat{\boldsymbol{s}},\nu)\rangle$ to distinguish between the EoS models. 

However, radio interferometer arrays are usually not sensitive to the sky-averaged spectrum \citep[see exceptions in][]{thy15a,pre15}, but are sensitive to the fluctuations instead, 
\begin{align}\label{eqn:dT_obs}
\delta T_\textrm{obs}(\hat{\boldsymbol{s}},\nu) &= \delta T_\textrm{b}(\hat{\boldsymbol{s}},\nu) - \langle \delta T_\textrm{b}(\hat{\boldsymbol{s}},\nu)\rangle.    
\end{align}
These fluctuations, for an arbitrary sightline $\hat{\boldsymbol{s}}$, are illustrated in the right-hand panels of Figure~\ref{fig:predicted_dTb}. The differences  between the two models at the onset, the end, the structures, and their scales are clearly noted. Many interferometer experiments with the HERA, LOFAR, MWA, LWA, and the SKA telescopes are underway to detect such fluctuations. 

\subsection{Compact Background Radio Sources} \label{sec:background}

This paper is focused on the detection of structures in the IGM by detecting the absorption of background radiation by H~{\sc i} atoms in the intervening IGM. Here, it is assumed that a compact background source of radiation other than the CMB is present at sufficiently high redshift. The background radiation source is assumed to be compact enough to have dimensions smaller than the transverse dimensions of an independent pixel (10\arcsec~ in this case; see below) in the synthesized image such that it appears as a point source. Although a quasar or AGN may serve as a typical example of a background source of radiation, the arguments can be extended to other sources such as radio afterglows from GRBs and star-forming radio galaxies as well. 

The background radiation model consists of a hypothetical radio emitter at a sufficiently high redshift ($z_\gamma$) to act as a background source of radiation with a continuum at low radio frequencies. Such sources are placed randomly at high redshifts without any spatial clustering. The intervening IGM is assumed to be at a redshift $z < z_\gamma$ not in the near-zone of the background radiation source. Therefore, it is still the global cosmological and astrophysical parameters and not the background radiation source that determine properties such as $T_\textrm{s}(\hat{\boldsymbol{s}},\nu)$ and $\tau_\nu(\hat{\boldsymbol{s}},\nu)$. Thus, only the $T_\gamma(\hat{\boldsymbol{s}},\nu,z)$ term in Equation~(\ref{eqn:Tb_obs}) is affected because of this change. Then, 
\begin{align}\label{eqn:T_rad}
    T_\gamma(\hat{\boldsymbol{s}},\nu,z=0) &= T_\textrm{rad}(\hat{\boldsymbol{s}},\nu,z=0) + T_\textrm{CMB}(z=0),
\end{align}
where $T_\textrm{rad}(\hat{\boldsymbol{s}},\nu,z=0)$ denotes the observed brightness temperature of the background radiation source. Brightness temperatures can be equivalently represented as a flux density by assuming an angular size ($\Omega$) for the pixel of interest ($\hat{\boldsymbol{s}}$) in the image using $S/\Omega = 2k_\textrm{B}\,(\nu/c)^2\,T$, where $k_\textrm{B}$ is the Boltzmann constant.

The flux density of the background radiation source in the observed frame is modeled as $S_\textrm{obs}^\textrm{rad}(\nu) = S_{150}(\nu/\nu_{150})^\alpha$, where the spectral index is assumed to be typically equal to that of Cygnus A, $\alpha=-1.05$ \citep{car02}. Three cases of $S_{150}$ are considered, namely, 1~mJy, 10~mJy, and 100~mJy at  $\nu_{150}=150$~MHz. These choices for $S_{150}$ are justified in \S\ref{sec:direct-detection}.

\subsection{Instrument}

Generic instrument model parameters are assumed to keep the discussion applicable to a wide range of modern radio interferometer arrays. The full width at half maximum (FWHM) of the synthesized PSF is chosen to be $\theta_\textrm{S}=10\arcsec$, which will be achievable with the SKA, the LOFAR, and the proposed expanded-GMRT \citep[EGMRT][]{patra19}. It is assumed that the integration time on each target background source is $\delta t=10$~hr. 

The antenna sensitivity is parameterized by $A_\textrm{e}/T_\textrm{sys}$, where $A_\textrm{e}$ is the effective area of an antenna, and $T_\textrm{sys}$ is the system temperature. Then the array sensitivity is expressed as  $N_\textrm{a}A_\textrm{e}/T_\textrm{sys}$, where $N_\textrm{a}$ is the number of antennas. Although $N_\textrm{a}A_\textrm{e}/T_\textrm{sys}$ will generally depend on the observing frequency, here it is assumed to be constant across the spectral passband. The $N_\textrm{a}A_\textrm{e}/T_\textrm{sys}$ for LOFAR, uGMRT, and SKA2 are assumed to be 100~m$^2$~K$^{-1}$, 70~m$^2$~K$^{-1}$, and 4000~m$^2$~K$^{-1}$, respectively \citep{braun19}. From \citet{patra19}, the EGMRT is projected to be approximately three times as sensitive as the uGMRT, and therefore, the $N_\textrm{a}A_\textrm{e}/T_\textrm{sys}$ for the EGMRT is assumed to be 210~m$^2$~K$^{-1}$. The SKALA-4.1 version of the aperture array system for the SKA1-low is projected to have $N_\textrm{a}A_\textrm{e}/T_\textrm{sys}=800$~m$^2$~K$^{-1}$ \citep{del18}. The $N_\textrm{a}A_\textrm{e}/T_\textrm{sys}$ for the SKA1-low will be used as a reference for comparison. 

The \emph{System Equivalent Flux Density} (SEFD) of the antenna can be written as $\textrm{SEFD} = 2k_\textrm{B}/(A_\textrm{e}/T_\textrm{sys})$. The noise flux density in each independent pixel (synthesized beam area) and each frequency channel of width $\delta\nu$ of the synthesized image cube over a duration of $\delta t$ is assumed to be drawn from a zero-mean Gaussian distribution whose standard deviation is given by \citep{TMS17,tay99}:
\begin{align}\label{eqn:noise-rms-theory}
    \delta S^\textrm{N} &= \frac{\textrm{SEFD}}{\eta\sqrt{n_\textrm{p}\,N_\textrm{b}\,(2\delta\nu\delta t)}} \approx \frac{\textrm{SEFD}}{\eta\,\frac{N_\textrm{a}}{\sqrt{2}}\sqrt{n_\textrm{p}\,(2\delta\nu\delta t)}}
\end{align}
where $2\delta\nu\delta t$ is the number of independent samples output by a complex correlator, $N_\textrm{b}=N_\textrm{a}(N_\textrm{a}-1)/2$ is the number of independent antenna spacings (baselines) in the interferometer array, $n_\textrm{p}$ is the number of orthogonal polarizations averaged, and $\eta$ is the overall system efficiency including loss in sensitivity due to the weighting schemes in synthesis imaging and data loss due to radio frequency interference (RFI), etc. The approximation in the last step results from assuming $N_\textrm{a}\gg 1$ and thus  $N_\textrm{b}\approx N_\textrm{a}^2/2$. Hereafter, $n_\textrm{p}=2$ is assumed. Therefore, the noise {\it rms} in each voxel of the synthesized image is:
\begin{align}\label{eqn:noise-rms}
    \delta S^\textrm{N} &\approx \frac{2k_\textrm{B}}{\eta\left(\frac{N_\textrm{a}A_\textrm{e}}{T_\textrm{sys}}\right)\sqrt{2\delta\nu\delta t}}
\end{align}

Although the sensitivity, $N_\textrm{a}A_\textrm{e}/T_\textrm{sys}$, is assumed to remain constant within the passband, it will vary in practice. These nominal values adopted are approximately the average across the entire passband and were found to be $\approx 20$\% lower (higher) at lower (higher) frequencies relative to the center of the passband \citep{del18,braun19,patra19}. The effect of this systematic variation in the different redshift subbands will be discussed later. 

\section{Absorption of 21~cm Line against Compact Background Radio Sources} \label{sec:absorption}

Consider a hypothetical compact background source of radiation (e.g. AGN, star-forming galaxy, or a GRB afterglow in radio wavelengths) to be present along one or more arbitrarily selected sightlines. Along those directions, $\hat{\boldsymbol{s}}$, the net background radiation in the observed frame can be written from Equation~(\ref{eqn:T_rad}) as:
\begin{align}\label{eqn:S_rad}
    S_\gamma(\hat{\boldsymbol{s}},\nu,z=0) &= S_\textrm{obs}^\textrm{rad}(\hat{\boldsymbol{s}},\nu,z=0) + S_\textrm{CMB}(\nu,z=0).
\end{align}
The corresponding flux density enclosed in the pixel in the observed frame, from Equation~(\ref{eqn:Tb_obs}), is:
\begin{align}\label{eqn:S_net}
    S_\textrm{net}(\hat{\boldsymbol{s}},\nu) &\approx \left[S_\textrm{s}(\hat{\boldsymbol{s}},\nu,z=0)-S_\gamma(\hat{\boldsymbol{s}},\nu,z=0)\right]\tau_\nu(\hat{\boldsymbol{s}},\nu) \nonumber\\ 
    &\qquad\qquad\qquad + S_\gamma(\hat{\boldsymbol{s}},\nu,z=0),
\end{align}

\begin{figure}
\includegraphics[width=\linewidth]{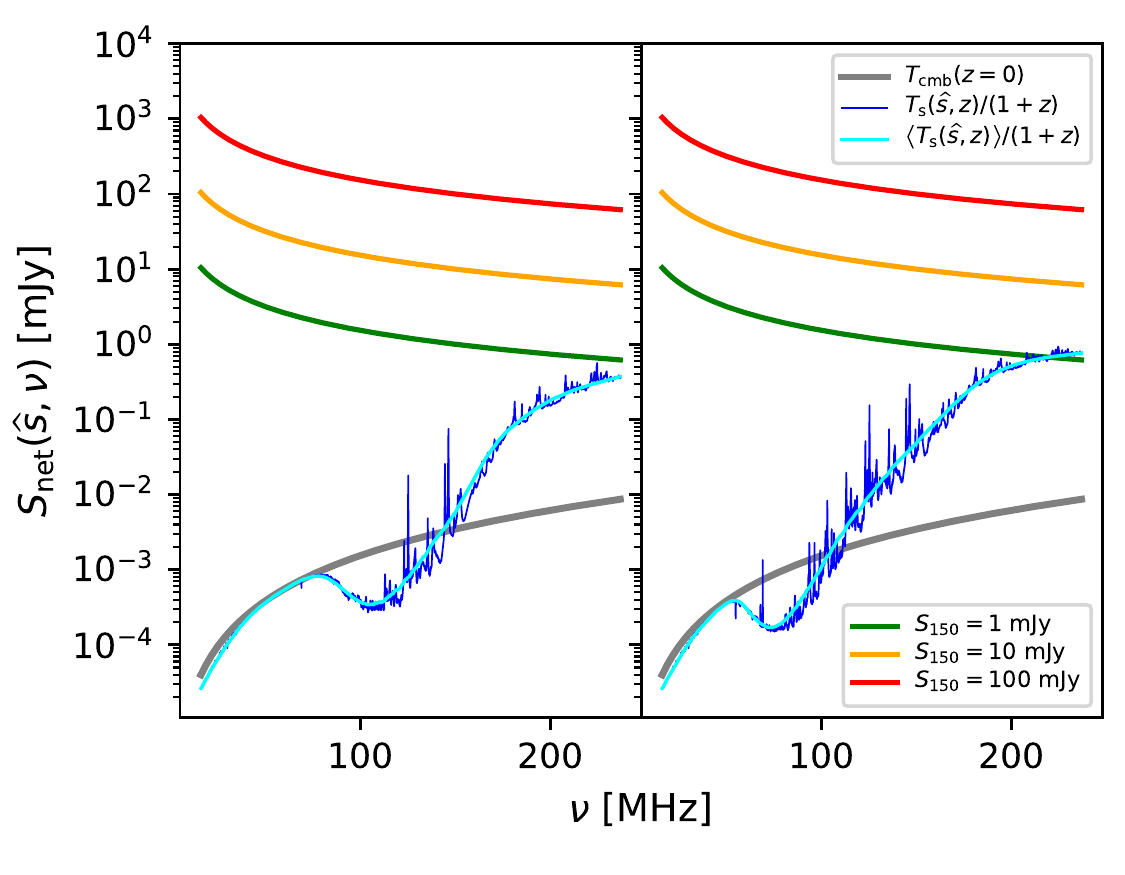}
\caption{{\it Left}: Different components of flux density in the observed frame contributing to the net observable flux density in the {\tt BRIGHT GALAXIES} model (EoS model 1). $T_\textrm{CMB}$ (gray), the spin temperature along an arbitrary direction (blue) and averaged over all directions (cyan), have been converted to flux densities in the observed frame assuming an angular resolution of $\theta_\textrm{s}=10\arcsec$ for the pixel. The continuum spectra (spectral index, $\alpha=-1.05$) of compact background radiation sources parameterized by the flux density at 150~MHz, $S_{150}=1$~mJy, 10~mJy, and 100~mJy are shown in green, orange, and red, respectively. {\it Right}: Same as the left but for the {\tt FAINT GALAXIES} model (EoS model 2). 
\label{fig:fluxdensity_comparison}}
\end{figure}

Figure~\ref{fig:fluxdensity_comparison} shows the different components of flux density contributing to  $S_\textrm{net}(\hat{\boldsymbol{s}},\nu)$ in Equation~(\ref{eqn:S_net}) for the EoS models 1 (left) and 2 (right). Varying levels of flux density from a hypothetical compact background object at a sufficiently high redshift are shown in green ($S_{150}=1$~mJy), orange ($S_{150}=10$~mJy), and red ($S_{150}=100$~mJy) in the observed frame. The gray curve shows $T_\textrm{CMB}(z=0)$ converted to flux density in the selected pixel for reference. The blue curve shows the $T_\textrm{s}(\hat{\boldsymbol{s}},\nu)/(1+z)$ (spin temperature spectrum in the chosen direction in the observed frame) converted to flux density in the selected pixel, while the cyan curve shows $\langle T_\textrm{s}(\hat{\boldsymbol{s}},\nu)\rangle/(1+z)$ converted to flux density units. The spectral index, $\alpha$, characterizing the continuum spectrum from the compact background object increases the flux density toward lower frequencies, whereas the CMB flux density in the pixel decreases as $\nu^{-2}$ in the Rayleigh-Jeans approximation. It is noted that even $\lesssim 1$~mJy radiation sources can be significantly stronger sources of background radiation, by a few orders of magnitude, relative to the CMB at the frequencies considered here.

Because an interferometer array will usually not be sensitive to the sky-averaged signal, the observed flux density will be missing this sky-averaged component. If $\langle \delta S(\hat{\boldsymbol{s}},\nu)\rangle$ corresponds to $\langle \delta T_\textrm{b}(\hat{\boldsymbol{s}},\nu)\rangle$ in Equation~(\ref{eqn:dT_obs}) obtained with only the CMB as the background radiation, then the interferometer will measure:
\begin{align}\label{eqn:S_obs}
    S_\textrm{obs}(\hat{\boldsymbol{s}},\nu) &\approx \left[S_\textrm{s}(\hat{\boldsymbol{s}},\nu,z=0)-S_\gamma(\hat{\boldsymbol{s}},\nu,z=0)\right]\tau_\nu(\hat{\boldsymbol{s}},\nu) \nonumber\\ 
    &\qquad\quad + S_\gamma(\hat{\boldsymbol{s}},\nu,z=0) + S_\textrm{fg}(\hat{\boldsymbol{s}},\nu,z=0) \nonumber \\
    &\qquad\quad - \langle\delta S(\hat{\boldsymbol{s}},\nu)\rangle,
\end{align}
where a foreground radiation term, $S_\textrm{fg}(\hat{\boldsymbol{s}},\nu,z=0)$, has been included for generality. $S_\textrm{fg}(\hat{\boldsymbol{s}},\nu,z=0)$ in turn includes the classical source confusion noise and the chromatic sidelobes of the residual sources of confusion. Then, the differential flux density observed by the interferometer with the foreground and background radiation (both compact background source and the CMB) removed in the selected pixel will be:
\begin{align}\label{eqn:dS_obs}
    \Delta S_\textrm{obs}(\hat{\boldsymbol{s}},\nu) &= \left[S_\textrm{s}(\hat{\boldsymbol{s}},\nu,z=0)-S_\gamma(\hat{\boldsymbol{s}},\nu,z=0)\right]\tau_\nu(\hat{\boldsymbol{s}},\nu) \nonumber \\
    &\qquad\quad - \langle\delta S(\hat{\boldsymbol{s}},\nu)\rangle.
\end{align}
Hereafter, the foreground contribution, $S_\textrm{fg}(\hat{\boldsymbol{s}},\nu,z=0)$, is assumed to have been perfectly removed and is not considered further until \S\ref{sec:sidelobes}. If $S_\gamma(\hat{\boldsymbol{s}},\nu,z=0) \gg S_\textrm{s}(\hat{\boldsymbol{s}},\nu,z=0)$, then Equation~(\ref{eqn:dS_obs}) reduces to:
\begin{align}\label{eqn:dS_obs_bright_bg}
    \Delta S_\textrm{obs}(\hat{\boldsymbol{s}},\nu) &\approx -S_\gamma(\hat{\boldsymbol{s}},\nu,z=0)\,\tau_\nu(\hat{\boldsymbol{s}},\nu) - \langle\delta S(\hat{\boldsymbol{s}},\nu)\rangle.
\end{align}
This is the usual scenario considered in approaches aiming to directly detect the absorption features in the spectrum \citep{car02,car04,car07,fur06a,mac12,cia13,cia15a,cia15b}. 

\begin{figure*}
\centering
% \subfloat[][EoS model 1 ({\tt BRIGHT GALAXIES})\label{fig:dS_obs_bright_gal}]{\includegraphics[width=0.47\textwidth]{figures/observable_fluxdensity_Bright_galaxies_fiducial_1024.pdf}}
% \subfloat[][EoS model 2 ({\tt FAINT GALAXIES})\label{fig:dS_obs_faint_gal}]{\includegraphics[width=0.47\textwidth]{figures/observable_fluxdensity_Faint_galaxies_fiducial_1024.pdf}}
\subfloat[][EoS model 1 ({\tt BRIGHT GALAXIES})\label{fig:dS_obs_bright_gal}]{\includegraphics[width=0.47\textwidth]{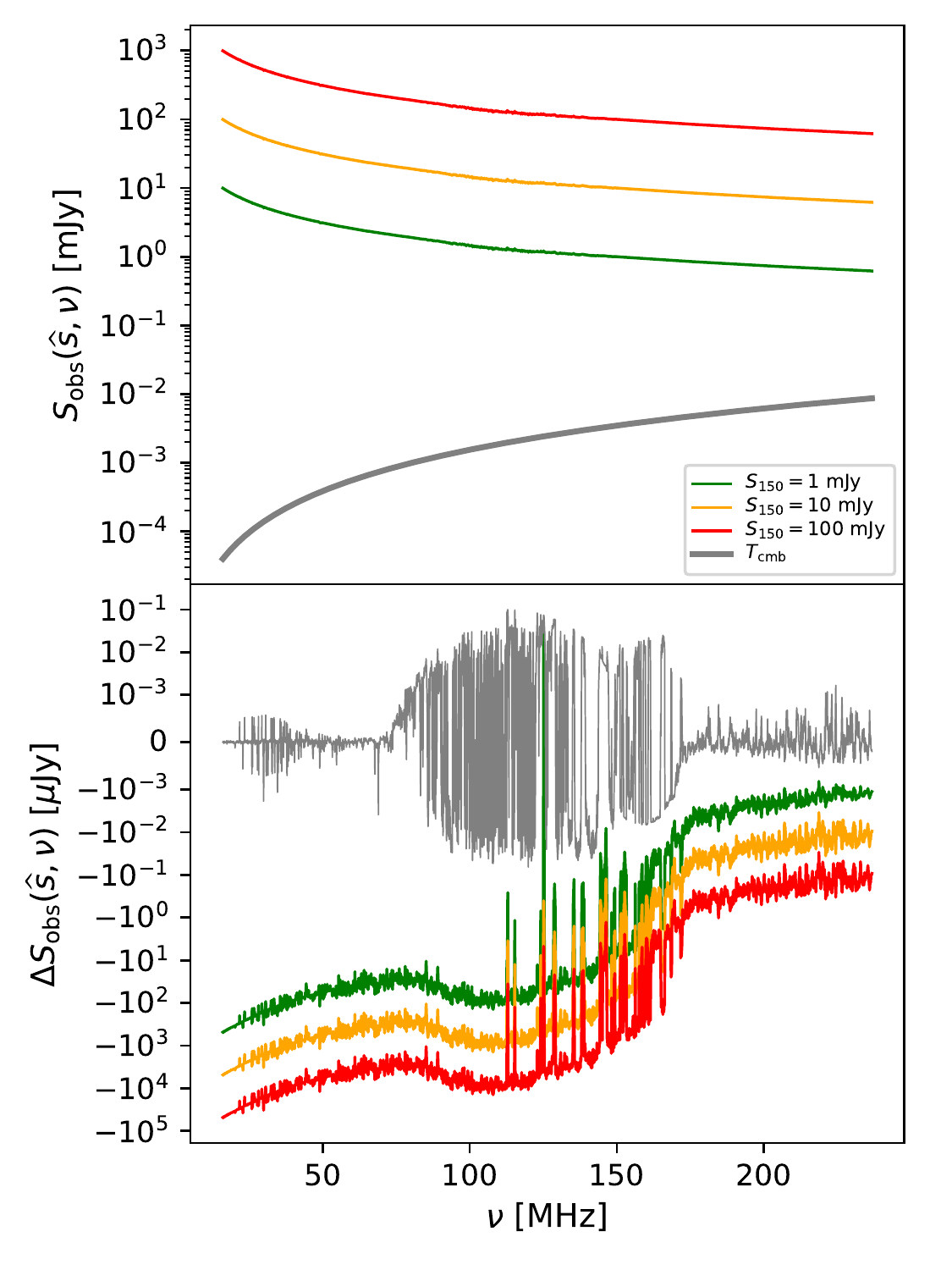}}
\subfloat[][EoS model 2 ({\tt FAINT GALAXIES})\label{fig:dS_obs_faint_gal}]{\includegraphics[width=0.47\textwidth]{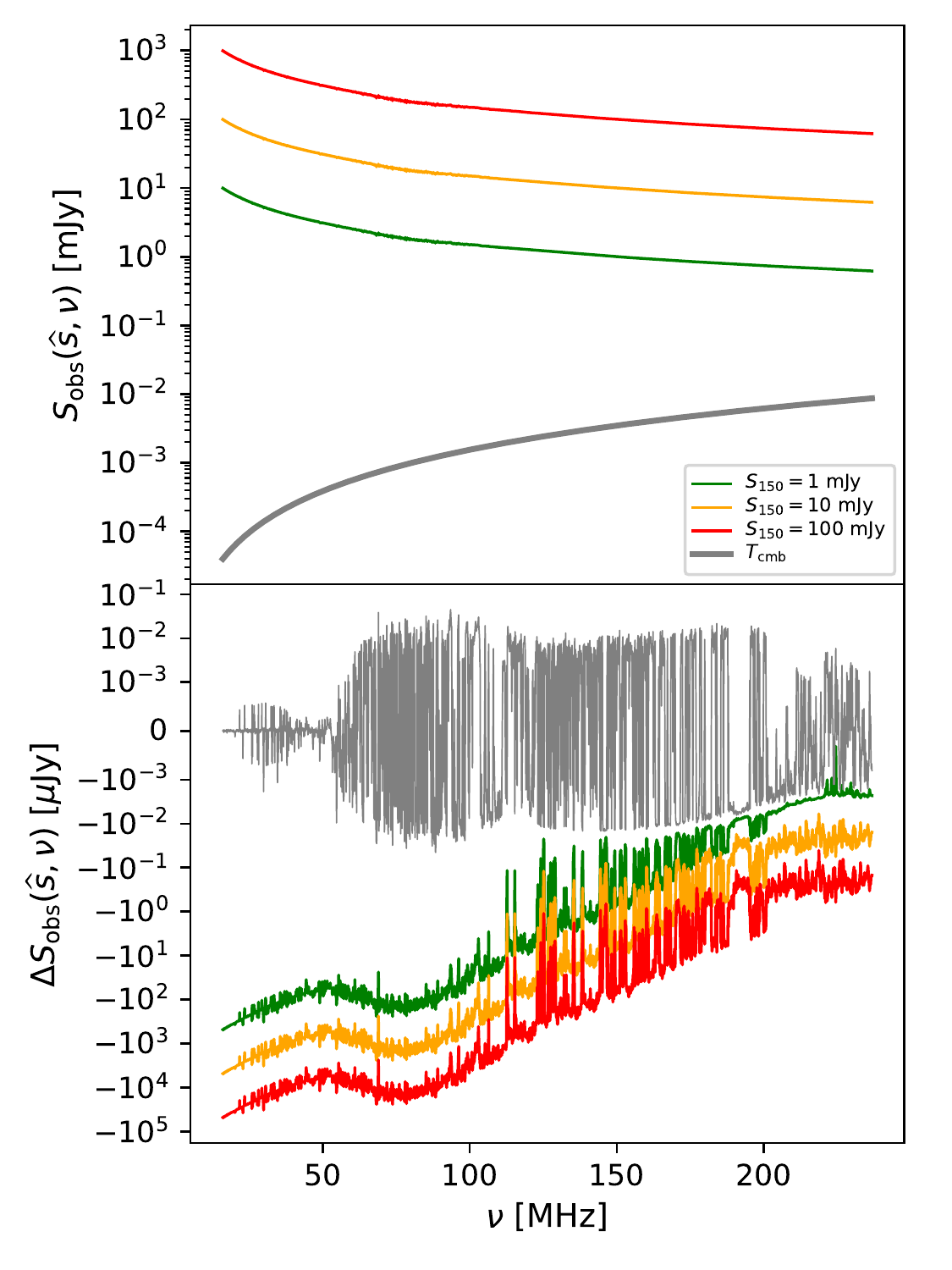}}
\caption{{\it Top}: Spectral evolution of the net flux densities in the observed frame measured by an interferometer. The compact background source of radiation is assumed to be hypothetically placed at a sufficiently high redshift well beyond the lowest frequency shown here. It is unrealistic but allows the expected absorption features to be visualized over the entire frequency range. The legend is the same as in Figure~\ref{fig:fluxdensity_comparison}. The absorption features in the continuum are barely visible in the spectra. The CMB flux density in the observed frame is shown for reference in gray. {\it Bottom}: Spectral evolution of the differential flux densities in the observed frame measurable by an interferometer after subtracting the background continuum, given by Equation~(\ref{eqn:dS_obs}). With only the CMB as the background radiation, the fluctuations oscillate to both positive and negative values. With a compact continuum source as background radiation placed at a sufficiently high redshift, the fluctuations are mostly negative (indicating absorption) and are many orders of magnitude stronger (even for $S_{150}=1$~mJy) than those with only the CMB as the background. Thus, a compact background radiation source provides a boost to the fluctuations (observed predominantly as absorption) in the spectrum. The $y$-axis is logarithmically scaled along both positive and negative directions. \label{fig:dS_obs}}
\end{figure*}

Figure~\ref{fig:dS_obs_bright_gal} shows the net flux density observed in the chosen pixel in the interferometer array image (top) and the differential flux density, $\Delta S_\textrm{obs}(\hat{\boldsymbol{s}},\nu)$ (bottom) in Equation~(\ref{eqn:dS_obs}) after subtracting the continuum flux density, $S_\gamma(\hat{\boldsymbol{s}},\nu,z=0)$, for EoS model 1. Figure~\ref{fig:dS_obs_faint_gal} is the same as Figure~\ref{fig:dS_obs_bright_gal} but for the EoS model 2. The differences in the signatures seen between the two EoS models in the bottom panels strongly correspond to the trends seen in the respective optical depths shown in Figure~\ref{fig:optical_depth}. Three prominent features are noted. First, as seen from Figure~\ref{fig:fluxdensity_comparison}, typically $S_\gamma(\hat{\boldsymbol{s}},\nu,z=0) > S_\textrm{s}(\hat{\boldsymbol{s}},\nu,z=0)$ even for the weakest value of $S_{150}=1$~mJy. Therefore, from Equation~(\ref{eqn:dS_obs}), the observed differential flux density, $\Delta S_\textrm{obs}(\hat{\boldsymbol{s}},\nu)<0$ typically and thus predominantly manifests as absorption for all values of $S_{150}$ used here. 
Second, the strength of absorption against the compact source of background radiation even as weak as $S_{150}=1$~mJy is several orders of magnitude stronger than that against the CMB background, resulting in a boost of the absorption signature. Thirdly, the absorption strength increases significantly at lower frequencies because it depends on the product of $S_\gamma(\hat{\boldsymbol{s}},\nu,z=0)$ and $\tau_\nu(\hat{\boldsymbol{s}},\nu)$ (from Equation~(\ref{eqn:S_obs}) and Equation~(\ref{eqn:dS_obs})), both of which increase significantly toward lower frequencies. 

Therefore, even radio-faint compact background objects can boost the absorption depth observed in the differential flux density, $\Delta S_\textrm{obs}(\hat{\boldsymbol{s}},\nu)$, by a few to several orders of magnitude relative to that against the CMB depending on the redshift. However, depending on the strength of the background radiation and the sensitivity of the observing instrument, that may still be insufficient for a direct detection in the spectra and will require statistical methods, which is addressed later. 

Figure~\ref{fig:dS_obs} was obtained by considering a compact radio background source hypothetically placed at a sufficiently high redshift with an observed flux density parameterized by $S_{150}$, with the sole purpose of being able to extend the analysis to any redshifted frequency subband available in the simulated 21cmFAST light-cone cubes. However, such a scenario of finding a compact background radiation source at any arbitrarily high redshift is unrealistic. In practice, the absorption features will be observed only at redshifts between the object and the observer, that is, at  frequencies higher (bluer) than that corresponding to the redshift of the background object.

The finite transverse resolution of an independent pixel ($\theta_\textrm{s}=10$\arcsec~ in this case) implies that any variations of $T_s$ on transverse scales finer than this resolution will become inaccessible. $\delta r_x = \delta r_y = 1.059375\,h^{-1}\,$cMpc corresponds to angular sizes of 35\farcs2 and 32\farcs7 at $z=8$ and $z=11$, respectively, for the adopted cosmological parameters. Thus, $\theta_\textrm{s}$ is smaller than the angular sizes of the voxel. And the 1D line-of-sight power spectrum approach considered here will only probe the line of sight for scales larger than $\delta r_z = 1.059375\,h^{-1}\,$cMpc. Therefore, the information on smaller scales in $T_s$ lost due to the transverse resolution will have no lossy effect on the line-of-sight scales probed by the 1D power spectrum.

\section{Detection methods} \label{sec:methods}

Three different redshifts, $z=8$, 9.5, and 11, are considered here for the detection of the absorption features in the spectra. These correspond to redshifted frequencies, $\nu_z\simeq 157.8$~MHz, 135.3~MHz, and 118.4~MHz, and wavelengths, $\lambda_z\simeq 1.9$~m, 2.2~m, and 2.5~m, respectively. $z=11$ is chosen to represent the epoch when heating begins to dominate in the EoS model 1, while $z=8$ is representative of an epoch when reionization is quite advanced in both the EoS models.
A coeval redshift range, $\Delta z$, is chosen using $\Delta z/z=0.067$ such that the cosmic evolution within this redshift range, $\Delta z$, is not significant \citep{bow06,mcq06}. This corresponds to comoving line-of-sight depths, $\Delta r_z \simeq 105.5\,h^{-1}\,$cMpc, 93.7$\,h^{-1}\,$~cMpc, and 78.2$\,h^{-1}\,$cMpc at redshifts of $z=8$, 9.5, and 11, respectively. Equivalently, the effective subband bandwidths are $\Delta\nu_z\simeq 9.14$~MHz, 6.98~MHz, and 4.73~MHz, respectively, obtained using \citep{mor04}:
\begin{align}\label{eqn:df-drz}
    \frac{\mathrm{d}\nu}{\mathrm{d}r_z} &= \frac{\nu_\textrm{r} H_0 E(z)}{c (1+z)^2},
\end{align}
where $E(z)\equiv [\Omega_m (1+z)^3 + \Omega_k (1+z)^2 + \Omega_\Lambda]^{1/2}$. 

Usually, a radio interferometric observation results in a spectral image cube where the spectral axis is uniformly divided in frequency. However, the data products used here are in the form of light-cone cubes, where the coordinate corresponding to the spectral axis is uniform in the line-of-sight comoving distance, $r_z$. Thus, the frequency channel width assumed corresponds to the comoving width of the voxel, $\delta r_z$, in the 21cmFAST light-cone cube and hence depends on the redshift (see Equation~(\ref{eqn:df-drz})). Although $\mathrm{d}\nu/\mathrm{d}r_z$ varies with redshift and hence, the observing frequency, its variation is however negligible within the narrow spectral subbands. Thus, $\delta\nu_z \simeq 91.8$~kHz, 78.9~kHz, and 64~kHz at $z=8$, 9.5, and 11, respectively, with $\delta r_z=1.059375\,h^{-1}\,$cMpc. $\delta\nu_z$ is assumed to remain constant within the respective spectral subbands. Note that $\delta\nu_z$ denotes the frequency channel width whereas $\Delta\nu_z$ denotes the effective subband bandwidth.

Nominal values for the various quantities that have been introduced already or will be defined subsequently are listed in Table~\ref{tab:nominal-values} for reference.

\begin{deluxetable}{lCcC}[htb]
\tablecaption{Nominal values of different quantities.\label{tab:nominal-values}}
\tablehead{\colhead{Quantity} & \colhead{Symbol} & \colhead{Value} & \colhead{Unit}}
\tabletypesize{\scriptsize}
\startdata
Redshift & z & 8, 9.5, 11 & - \\
Subband center\tablenotemark{a} & \nu_z & 157.8, 135.3 118.4 & \textrm{MHz} \\
Wavelength center\tablenotemark{a} & \lambda_z & 1.9, 2.2, 2.5 & \textrm{m} \\
Coeval redshift fraction & \Delta z/z & 0.067 & - \\
Comoving depth\tablenotemark{b} & \Delta r_z & 105.5, 93.7, 78.2 & h^{-1}\,\textrm{cMpc} \\
Comoving resolution\tablenotemark{c} & \delta r_x, \delta r_y, \delta r_z & 1.059375 & h^{-1}\,\textrm{cMpc} \\
Subband bandwidth\tablenotemark{d} & \Delta\nu_z & 9.14, 6.98, 4.73 & \textrm{MHz} \\
Spectral resolution\tablenotemark{d} & \delta\nu_z & 91.8, 78.9, 64 & \textrm{kHz} \\
Background sources\tablenotemark{e} & N_\gamma & 1 or 100 & - \\
Source flux density\tablenotemark{f} & S_{150} & 1, 10, 100 & \textrm{mJy} \\
Spectral index\tablenotemark{f} & \alpha & $-1.05$ & - \\
Integration time\tablenotemark{g} & \delta t & 10 & \textrm{hr} \\
Total time\tablenotemark{h} & \Delta t & $\le 1000$ & \textrm{hr} \\ 
Angular resolution\tablenotemark{i} & \theta_\textrm{s} & 10 & \textrm{arcsec} \\
Image pixel size & \Omega & $\pi\left(\frac{\theta_\textrm{s}}{2}\right)^2$ & \textrm{sr} \\
Field of view\tablenotemark{i} & \theta_\textrm{p} & 5 & \textrm{degree} \\
Array sensitivity\tablenotemark{j} & \frac{N_\textrm{a}A_\textrm{e}}{T_\textrm{sys}} & 800 & \textrm{m}^2\,\textrm{K}^{-1} \\
System efficiency & \eta & 1.0 & - \\
S/N detection threshold & \textrm{S/N} & 5 & - \\
\enddata
\tablenotetext{a}{$\nu_z = \nu_r/(1+z) = c/\lambda_z$}
\tablenotetext{b}{From cosmology equations for comoving line-of-sight distance.}
\tablenotetext{c}{From 21cmFAST simulations.}
\tablenotetext{d}{Derived from Equation~(\ref{eqn:df-drz}).}
\tablenotetext{e}{$N_\gamma=1$ for $P^\textrm{N}$ in Figure~\ref{fig:pspec_redshifts}, $N_\gamma=100$ in Equation~(\ref{eqn:Aeff_over_T}).}
\tablenotetext{f}{Defined at $\nu_{150}=150$~MHz, $S_\textrm{obs}^\textrm{rad}(\nu)=S_{150}(\nu/\nu_{150})^\alpha$. $\alpha$ chosen to match that of Cygnus~A \citep{car02}.}
\tablenotetext{g}{Defined per background source, $\delta t = 2\,\delta t_{0.5}$.}
\tablenotetext{h}{$\Delta t \le N_\gamma\,\delta t$, inequality applies when more than one background source lies in the same field of view.}
\tablenotetext{i}{FWHM of synthesized PSF and primary beam of antenna power pattern for $\theta_\textrm{s}$ and $\theta_\textrm{p}$, respectively, assumed to be independent of frequency.}
\tablenotetext{j}{Matches anticipated SKA1-low performance \citep{del18}, assumed to be independent of frequency.}
\end{deluxetable}

\subsection{Direct Detection of Absorption Spectra}\label{sec:direct-detection}

A `direct detection' approach will aim to detect absorption features directly in the spectrum along a selected pixel in the image that usually contains a compact source of bright ($\gtrsim 10$--100~mJy at 150~MHz) background radiation \citep{car02,car04,car07,fur06b,mac12,cia13,cia15a,cia15b}. The $S_{150}=100$~mJy case is included in this study to represent such a scenario. 

It is now becoming evident that there is a dearth of such bright radio quasars at high redshifts \citep{banados15,sax17,bolgar18}. However, in general, there is a growing list of high-redshift quasars \citep{banados16}\footnote{An updated list of high-redshift quasars can be found at \url{https://users.obs.carnegiescience.edu/~ebanados/high-z-qsos.html}}, many of which could be weak radio emitters and yet provide a statistically significant source of background radiation. It was noted in \citet{ewa14} that background quasars with flux densities $\sim 1$--10~mJy could contribute significantly to have an observable effect on the standard 3D power spectrum particularly on small scales. Hence, $S_{150}=1$~mJy and 10~mJy are also considered here for statistical purposes. The likelihood of the presence of such a faint radio population will be discussed later in the context of currently available models. 

\begin{figure*}
\centering
% \subfloat[][EoS model 1 ({\tt BRIGHT GALAXIES})\label{fig:dS_obs_redshifts_bright_gal}]{\includegraphics[width=\textwidth]{figures/observable_differential_fluxdensity_Bright_galaxies_fiducial_1024.pdf}} \\
% \subfloat[][EoS model 2 ({\tt FAINT GALAXIES})\label{fig:dS_obs_redshifts_faint_gal}]{\includegraphics[width=\textwidth]{figures/observable_differential_fluxdensity_Faint_galaxies_fiducial_1024.pdf}}
\subfloat[][EoS model 1 ({\tt BRIGHT GALAXIES})\label{fig:dS_obs_redshifts_bright_gal}]{\includegraphics[width=\textwidth]{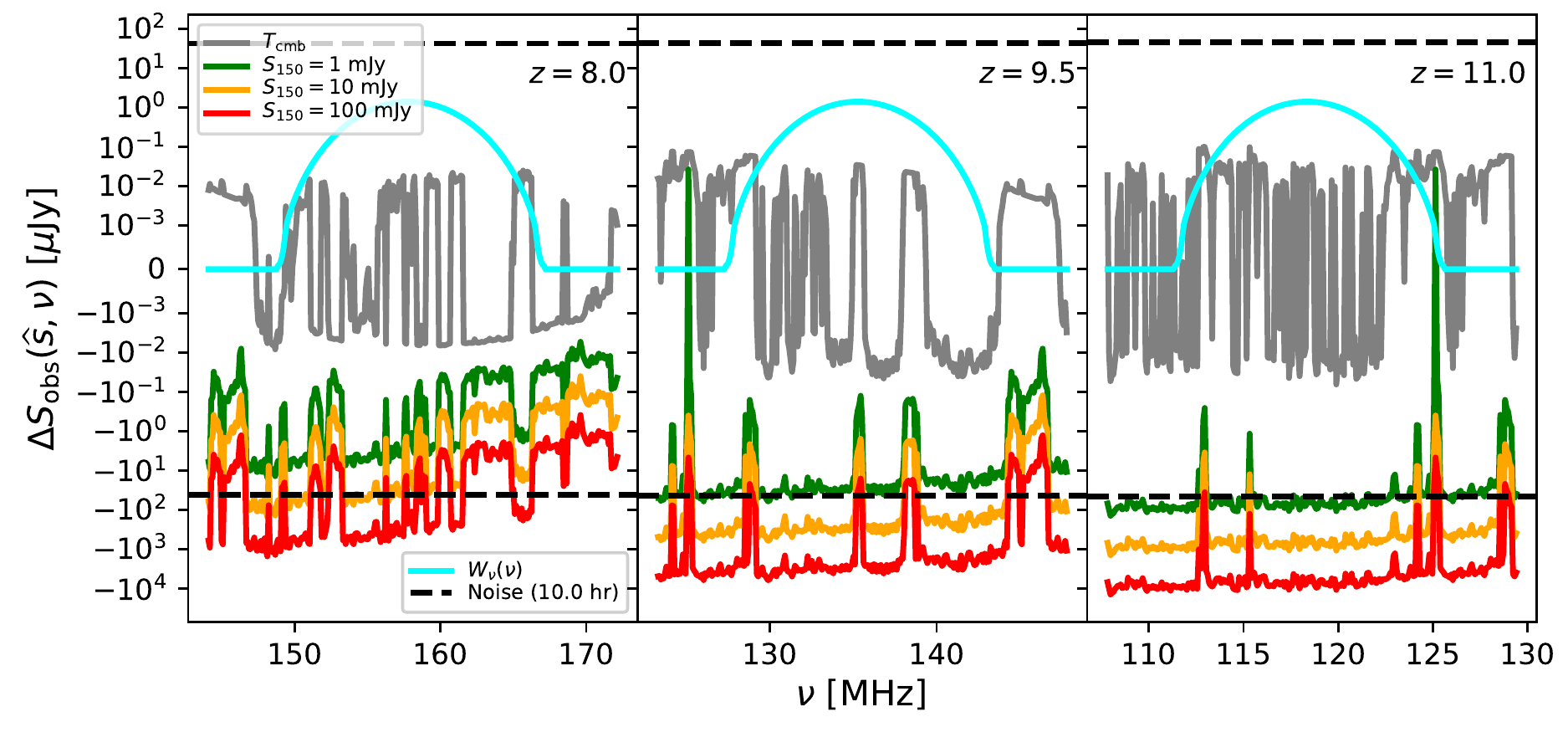}} \\
\subfloat[][EoS model 2 ({\tt FAINT GALAXIES})\label{fig:dS_obs_redshifts_faint_gal}]{\includegraphics[width=\textwidth]{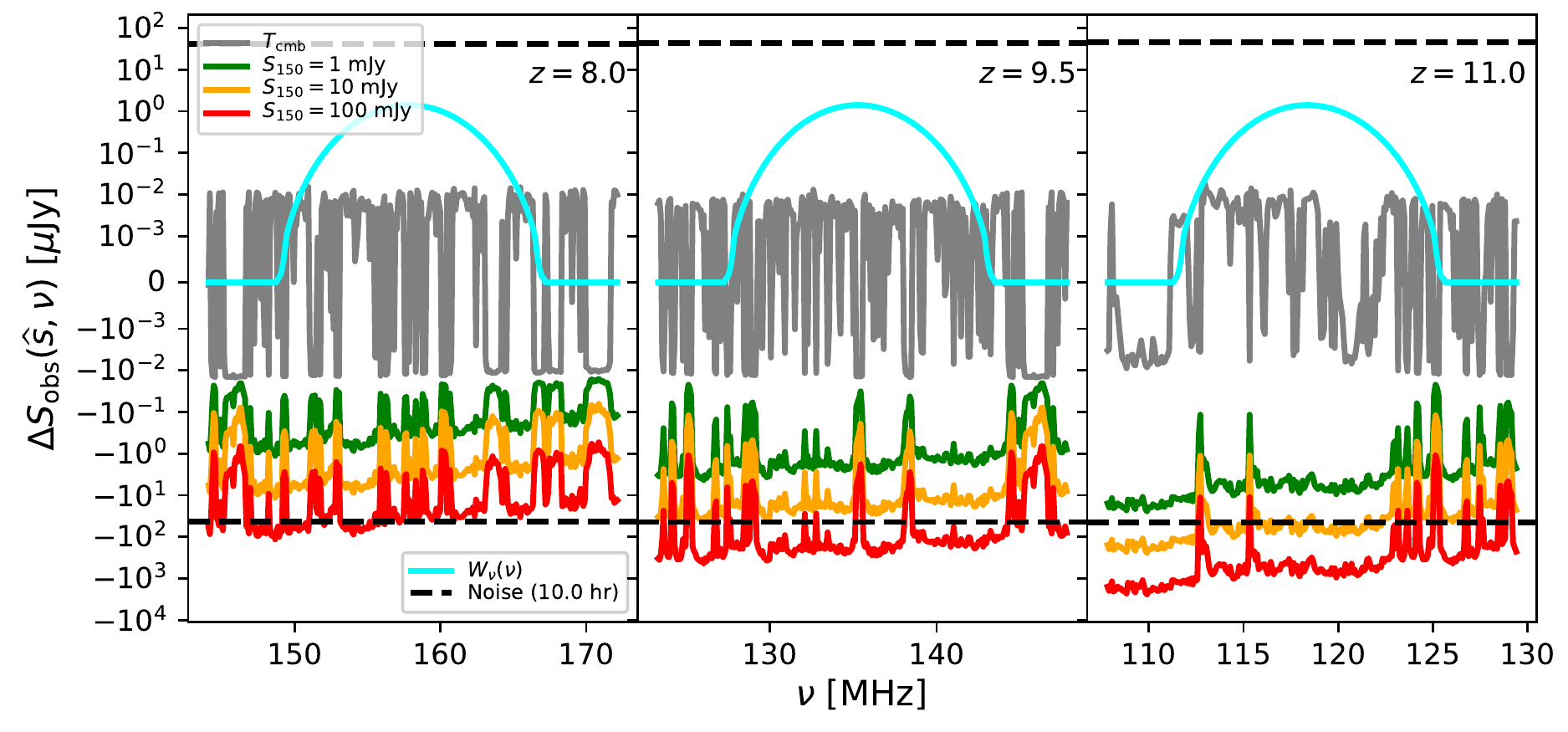}}
\caption{Same as the lower panels in Figure~\ref{fig:dS_obs} but restricted to the selected spectral subbands centered on $\nu_z=157.8$~MHz (left), 135.3~MHz (middle), and 118.4~MHz (right) corresponding to $z=8$, $z=9.5$, and $z=11$, respectively. The cyan curve shows $W_\nu(\nu)$, the spectral window function of effective bandwidth $\Delta\nu_z=9.14$~MHz, 6.98~MHz, and 4.73~MHz in the respective subbands, chosen here to be a {\it Blackman-Harris} window function \citep{har78} that will be applied to the differential flux density spectra while computing the 1D power spectrum. The black dashed lines show the {\it rms} of the zero-mean Gaussian noise of 42.4~$\mu$Jy, 43.8~$\mu$Jy, and 45.3~$\mu$Jy from Equation~(\ref{eqn:noise-rms}) in each image voxel with spectral channel width $\Delta\nu_z=91.8$~kHz, 78.9~kHz, and 64~kHz.in the respective subbands for efficiency factor $\eta=1$, array sensitivity parameter $N_\textrm{a}A_\textrm{e}/T_\textrm{sys}=800\,\textrm{m}^2\,\textrm{K}^{-1}$, and integration time $\delta t=10$~hr. The fluctuations (predominantly in absorption) against the compact background radiation sources considered here are at least a few orders of magnitude stronger than that against the CMB radiation. Generally, the amplitude of the absorption is larger with increasing redshift. The $y$-axis is logarithmically scaled along both positive and negative directions. \label{fig:dS_obs_redshifts}}
\end{figure*}

The differential flux densities expected from different sources of background radiation for EoS models 1 and 2 are shown in Figure~\ref{fig:dS_obs_redshifts_bright_gal} and Figure~\ref{fig:dS_obs_redshifts_faint_gal}, respectively. These are the same as in the lower panels of Figure~\ref{fig:dS_obs_bright_gal} and Figure~\ref{fig:dS_obs_faint_gal}, respectively, but the frequencies are restricted to the chosen subband. The cyan curve represents a spectral weighting \citep[Blackman-Harris window function;][]{har78} that is applied to the differential flux density spectra to compute the power spectrum and will be discussed in more detail below. The black dashed lines denote the thermal noise {\it rms} ($\pm 1$ standard deviation), $\delta S^\textrm{N}$, from a synthesized image produced over a duration of $\delta t=10$~hr for a channel width specified above for each of the subbands. From Equation~(\ref{eqn:noise-rms}), $\delta S^\textrm{N}=42.2\,\mu$Jy, $43.8\,\mu$Jy, and $45.3\,\mu$Jy for frequency subbands corresponding to $z=8$, 9.5, and 11 respectively, assuming the array sensitivity factor, $N_\textrm{a}A_\textrm{e}/T_\textrm{sys}=800$~m$^2$~K$^{-1}$, and efficiency factor, $\eta=1$. In practice,  $N_\textrm{a}A_\textrm{e}/T_\textrm{sys}$ will not remain constant in the subband. If its variation is nominally assumed to be 20\% lower and higher at, respectively, the lowest and the highest frequency subbands considered here, the corresponding noise levels will also worsen and improve by the respective amounts roughly because of its  $\sim (N_\textrm{a}A_\textrm{e}/T_\textrm{sys})^{-1}$ dependence and will not significantly affect the conclusions for a direct detection approach.

Figure~\ref{fig:dS_obs_redshifts_bright_gal} illustrates that in the EoS model 1, at $z=8$, most of the absorption features at $\nu\lesssim 160$~MHz against the $S_{150}=100$~mJy compact source are detectable with $\gtrsim 5\sigma$ significance. However, none of the absorption features against the weaker background sources is detectable with high signal-to-noise ratio (S/N). At higher redshifts, the absorption features are deeper and hence most of those even against the $S_{150}=10$~mJy source become detectable, while those against the $S_{150}=1$~mJy source are not detected at all at any of the redshifts considered. On the other hand, due to differences in optical depth and the timing of the reionization process, the absorption features in the EoS model 2 in Figure~\ref{fig:dS_obs_redshifts_faint_gal} are overall fainter relative to those in EoS model 1 and hence remain mostly undetectable or only marginally detectable except against the $S_{150}=100$~mJy background radiation at $z=9.5$ and $z=11$. Although the cosmic reionization models and the observing and instrument parameters ($\delta\nu \le 20$~kHz, $\delta t=1000$~hr, $N_\textrm{a}A_\textrm{e}/T_\textrm{sys}=1000$~m$^2$~K$^{-1}$ for SKA1-low) used were different in \citet{cia13,cia15a,cia15b}, the findings in the $S_{150}=100$~mJy case here are roughly consistent from the point of view of direct detection for the observing parameters considered here. 

Assuming statistical isotropy, the use of a statistical method such as a power spectrum can help improve sensitivity further and potentially detect more of the undetected or marginally detected features using a much larger number of sources of background radiation, including fainter objects. Unlike direct detection, the power spectrum is a statistical tool and cannot determine the exact location of the features or properties at any arbitrary location along the sightline. However, it can provide the distribution of power in the H~{\sc i} structures on different scales and is less susceptible to cosmic variance compared to a direct detection technique. Motivated by these merits, the 1D line-of-sight power spectrum technique is explored below. 

\subsection{Variance Statistic} \label{sec:var-stats}

The scale-dependent variance statistic of the differential flux density along a selected $\hat{\boldsymbol{s}}$, represented by a 1D line-of-sight power spectrum, is considered here.

\subsubsection{1D Power Spectrum along the Line of Sight} \label{sec:pspec}

The 1D power spectrum of the differential flux density along the sightline is considered along directions that contain a compact source of background radiation. Because $\nu$ and $r_z$ are closely related to each other as given by Equation~(\ref{eqn:df-drz}), $\Delta S_\textrm{obs}(\hat{\boldsymbol{s}},\nu)$ is equivalently expressible as $\Delta S_\textrm{obs}(\hat{\boldsymbol{s}},\nu) \equiv \Delta S^\prime_\textrm{obs}(\hat{\boldsymbol{s}},r_z)$. The {\it Fourier} transform of $\Delta S^\prime_\textrm{obs}(\hat{\boldsymbol{s}},r_z)$ is given by:
\begin{align}\label{eqn:comoving-FT}
    \Delta\widetilde{S}^\prime(\hat{\boldsymbol{s}},k_\parallel) &= \int W_{r_z}(r_z)\,\Delta S^\prime_\textrm{obs}(\hat{\boldsymbol{s}},r_z)\,e^{-ik_\parallel r_z}\,\mathrm{d}r_z,
\end{align}
where $W_{r_z}(r_z)$ is a window function applied along the $r_z$ coordinate, with an effective comoving depth of $\Delta r_z$. Its choice is usually influenced by the suppression of sidelobes and a high dynamic range \citep{thy13,thy15a,thy16,ved12} in the resulting {\it Fourier} transform. 

The 1D power spectrum along the line-of-sight coordinate, $r_z$, can be defined, analogous to the 3D power spectrum \citep{mor04,mcq06,par12a}, as:
\begin{align}\label{eqn:comoving-PS}
    P(\hat{\boldsymbol{s}},k_\parallel) &= \left|\Delta\widetilde{S}^\prime(\hat{\boldsymbol{s}},k_\parallel)\right|^2 \left(\frac{1}{\Delta r_z}\right)\left(\frac{\lambda_z^2}{2k_\textrm{B}\Omega}\right)^2.
\end{align}
The term $1/\Delta r_z$ is the normalization to account for the window function, $W_{r_z}(r_z)$. The last term $\lambda_z^2/(2k_\textrm{B}\Omega)$ converts flux density to equivalent temperature using the Rayleigh-Jeans law. In this paper, $P(\hat{\boldsymbol{s}},k_\parallel)$ is expressed in units of K$^2\,h^{-1}$cMpc. This derivation is appropriate from a theoretical viewpoint where the light-cone cubes are available in comoving coordinates. 

Below is an alternate but equivalent derivation based on an observational viewpoint, where the line-of-sight coordinate is represented by $\nu$. Analogous to the equations above, the {\it Fourier} transform of $\Delta S_\textrm{obs}(\hat{\boldsymbol{s}},\nu)$ is:
\begin{align}\label{eqn:spectral-FT}
    \Delta\widetilde{S}(\hat{\boldsymbol{s}},\xi) &= \int W_\nu(\nu)\,\Delta S_\textrm{obs}(\hat{\boldsymbol{s}},\nu)\,e^{-i2\pi\nu\xi}\,\mathrm{d}\nu,
\end{align}
where $W_\nu(\nu)\equiv W_{r_z}(r_z)$, but applied along the $\nu$ axis, with an effective bandwidth of $\Delta\nu_z$ corresponding to $\Delta r_z$ (Equation~(\ref{eqn:df-drz})). In this paper, $W_{r_z}(r_z)$ is a {\it Blackman-Harris} function \citet{har78} and is shown as the cyan curves in Figure~\ref{fig:dS_obs_redshifts}. Equation~(\ref{eqn:df-drz}) can be extended to \citep{mor04}:
\begin{align}\label{eqn:dkprll-dtau}
    \frac{1}{2\pi} \frac{\mathrm{d}k_\parallel}{\mathrm{d}\xi} = \frac{\mathrm{d}\nu}{\mathrm{d}r_z},
\end{align}
and the corresponding the 1D power spectrum defined in $\xi$-coordinates is \citep[derived analogously to 3D power spectrum in][]{mor04,mcq06,par12a}:
\begin{align}\label{eqn:comoving-PS-xi}
    P_\xi(\hat{\boldsymbol{s}},\xi) &= |\Delta\widetilde{S}(\hat{\boldsymbol{s}},\xi)|^2 \left(\frac{1}{\Delta\nu_z }\right) \left(\frac{\lambda_z^2}{2k_\textrm{B}\Omega}\right)^2, 
\end{align}
where $1/\Delta\nu_z$ is the normalization to account for the window function, $W_\nu(\nu)$. Also,
\begin{align}
    \frac{1}{2\pi}P(\hat{\boldsymbol{s}},k_\parallel)\mathrm{d}k_\parallel &= P_\xi(\hat{\boldsymbol{s}},\xi)\mathrm{d}\xi,
\end{align}
where the factor $2\pi$ arises from the {\it Fourier} transform convention used. Hence, 
\begin{align}\label{eqn:spectral-PS}
    P(\hat{\boldsymbol{s}},k_\parallel) &\approx \left|\Delta\widetilde{S}(\hat{\boldsymbol{s}},\xi)\right|^2 \left(\frac{1}{\Delta\nu_z}\right) \left(\frac{\Delta r_z}{\Delta\nu_z}\right) \left(\frac{\lambda_z^2}{2k_\textrm{B}\Omega}\right)^2,
\end{align}
where the factor $\Delta\nu_z/\Delta r_z$ is an approximation for the {\it Jacobian} in Equation~(\ref{eqn:dkprll-dtau}) assuming it does not vary significantly with frequency (or redshift) within the redshift subband. Equation~(\ref{eqn:spectral-PS}) is usually applicable from an observational viewpoint, where the image cube is available in which the line-of-sight dimension is uniformly sampled in $\nu$ rather than in $r_z$ but is an approximation due to the aforementioned reason.  

By using $\Delta\widetilde{S}(\hat{\boldsymbol{s}},k_\parallel) = (\Delta r_z/\Delta\nu_z)\Delta\widetilde{S}(\hat{\boldsymbol{s}},\xi)$, it can be verified that Equation~(\ref{eqn:comoving-PS}) and Equation~(\ref{eqn:spectral-PS}) are equivalent. In this paper, $P(\hat{\boldsymbol{s}},k_\parallel)$ was estimated using Equation~(\ref{eqn:comoving-PS}) to avoid potential inaccuracies resulting from interpolating the light-cone cube in comoving coordinates to spectral coordinates and due to the inherent approximation described above. Assuming statistical isotropy, the population mean of the underlying power spectra marginalized over directions, $\hat{\boldsymbol{s}}$, is denoted by $P(k_\parallel)\equiv\left\langle P(\hat{\boldsymbol{s}},k_\parallel)\right\rangle$.

\subsubsection{Thermal Noise Uncertainty in 1D Power Spectrum}\label{sec:pspec-noise}

In order to avoid potential noise bias in the positive-valued auto-power spectrum, it is assumed that the observing time available along each $\hat{\boldsymbol{s}}$ is divided into two halves ($\delta t_{0.5} = 0.5\,\delta t$) and imaged separately. This will result in a slightly higher image rms in each half, $\delta S_{0.5}^\textrm{N}=\sqrt{2}\,\delta S^\textrm{N}$. And the spectra from each half is {\it Fourier}-transformed from which the cross-power is computed. Such a cross-power spectrum will not be biased but will have a noise power {\it rms} a factor of 2 higher than that obtained from a completely coherently averaged image. 

From Equation~(\ref{eqn:spectral-FT}), the noise {\it rms} (including contributions from the real and imaginary parts) in each of the independent modes in the {\it Fourier} transform will be \citep{mor04,mcq06}: 
\begin{align}\label{eqn:noise-dspec-rms}
    \delta \widetilde{S}_{0.5}^\textrm{N} &= \frac{2k_\textrm{B}}{\eta N_\textrm{a}A_\textrm{e}/T_\textrm{sys}} \frac{1}{\sqrt{2\delta\nu_z\,\delta t_{0.5}}} \left(\frac{\Delta\nu_z}{\delta\nu_z}\right)^{1/2}\delta\nu_z \nonumber \\
    &= \frac{2k_\textrm{B}}{\eta N_\textrm{a}A_\textrm{e}/T_\textrm{sys}} \left(\frac{\Delta\nu_z}{2\,\delta t_{0.5}}\right)^{1/2}.
\end{align}
The cross-product of the {\it Fourier} transforms from the two halves of the measurements after converting flux density to equivalent temperature will have a standard deviation (including contributions from real and imaginary parts) in each of the {\it Fourier} modes:
\begin{align}
    C_{(\frac{1}{2},\frac{2}{2})}^\textrm{N} &= \left(\frac{\lambda_z^2}{2k_\textrm{B}\Omega}\,\delta \widetilde{S}_{0.5}^\textrm{N}\right)^2 = \left(\frac{\lambda_z^2 / \Omega}{\eta N_\textrm{a}A_\textrm{e}/T_\textrm{sys}}\right)^2 \frac{\Delta\nu_z}{2\,\delta t_{0.5}},
\end{align}
which has units of K$^2$~Hz$^2$. Therefore, the standard deviation of the noise power spectrum in each of the $k_\parallel$-modes (including contributions from both real and imaginary parts) is derived as in \S\ref{sec:pspec}:
\begin{align}\label{eqn:cross-power-noise-rms}
    P_{(\frac{1}{2},\frac{2}{2})}^\textrm{N} &= \left(\frac{\lambda_z^2 / \Omega}{\eta N_\textrm{a}A_\textrm{e}/T_\textrm{sys}}\right)^2\left(\frac{\Delta r_z}{2\Delta\nu_z\,\delta t_{0.5}}\right),
\end{align}
which is expressed in units of K$^2\,h^{-1}$cMpc. It can be easily verified by substituting $\delta t$ for $\delta t_{0.5}$ in Equation~(\ref{eqn:cross-power-noise-rms}) which in the case of a completely coherently synthesized image cube, the standard deviation of the so-obtained auto-noise power will be $P^\textrm{N}_{(\frac{1}{1},\frac{1}{1})} = (1/2)\,P_{(\frac{1}{2},\frac{2}{2})}^\textrm{N}$ (twice as sensitive) but will also suffer from a positive noise bias. In this paper, the standard deviation of the cross-power of noise is employed. 

Finally, by assuming statistical isotropy, additional sensitivity via reduction in sample variance and noise power can be obtained by the incoherent averaging of the 1D power over a number of independent directions, each of which contains a compact source of background radiation. Thus, $N_\gamma\equiv N_\gamma(z)$ is defined as the number of such compact background radiation sources available for observations at redshifts $> z$. Such an incoherent averaging in power spectrum will yield improvements in sensitivity by a factor of $\sqrt{N_\gamma}$. Thus, the final noise standard deviation in the 1D power spectrum will be:
\begin{align}\label{eqn:cross-power-noise-rms-avg}
    P^\textrm{N} &= \left\langle\left(P_{(\frac{1}{2},\frac{2}{2})}^\textrm{N}\right)^2\right\rangle^{1/2} =  \left[\frac{1}{N_\gamma}\,\left(P_{(\frac{1}{2},\frac{2}{2})}^\textrm{N}\right)^2\right]^{1/2} \nonumber \\
    &= \frac{1}{\sqrt{N_\gamma}}\left(\frac{\lambda_z^2/\Omega}{\eta N_\textrm{a}A_\textrm{e}/T_\textrm{sys}}\right)^2\left(\frac{\Delta r_z}{2\Delta\nu_z\,\delta t_{0.5}}\right).
\end{align}
This noise estimate has been verified against simulations of a sufficiently large number of random realizations of noise spectra based on Equation~(\ref{eqn:noise-rms}) and propagating them through the power spectrum methodology. 

\begin{figure*}
\centering
% \subfloat[][EoS model 1 (\tt{BRIGHT GALAXIES})\label{fig:pspec_bright_gal}]{\includegraphics[width=\textwidth]{figures/Pspec_pre_post_modification_Bright_galaxies_fiducial_1024.pdf}} \\
% \subfloat[][EoS model 2 (\tt{FAINT GALAXIES})\label{fig:pspec_faint_gal}]{\includegraphics[width=\textwidth]{figures/Pspec_pre_post_modification_Faint_galaxies_fiducial_1024.pdf}}
\subfloat[][EoS model 1 (\tt{BRIGHT GALAXIES})\label{fig:pspec_bright_gal}]{\includegraphics[width=\textwidth]{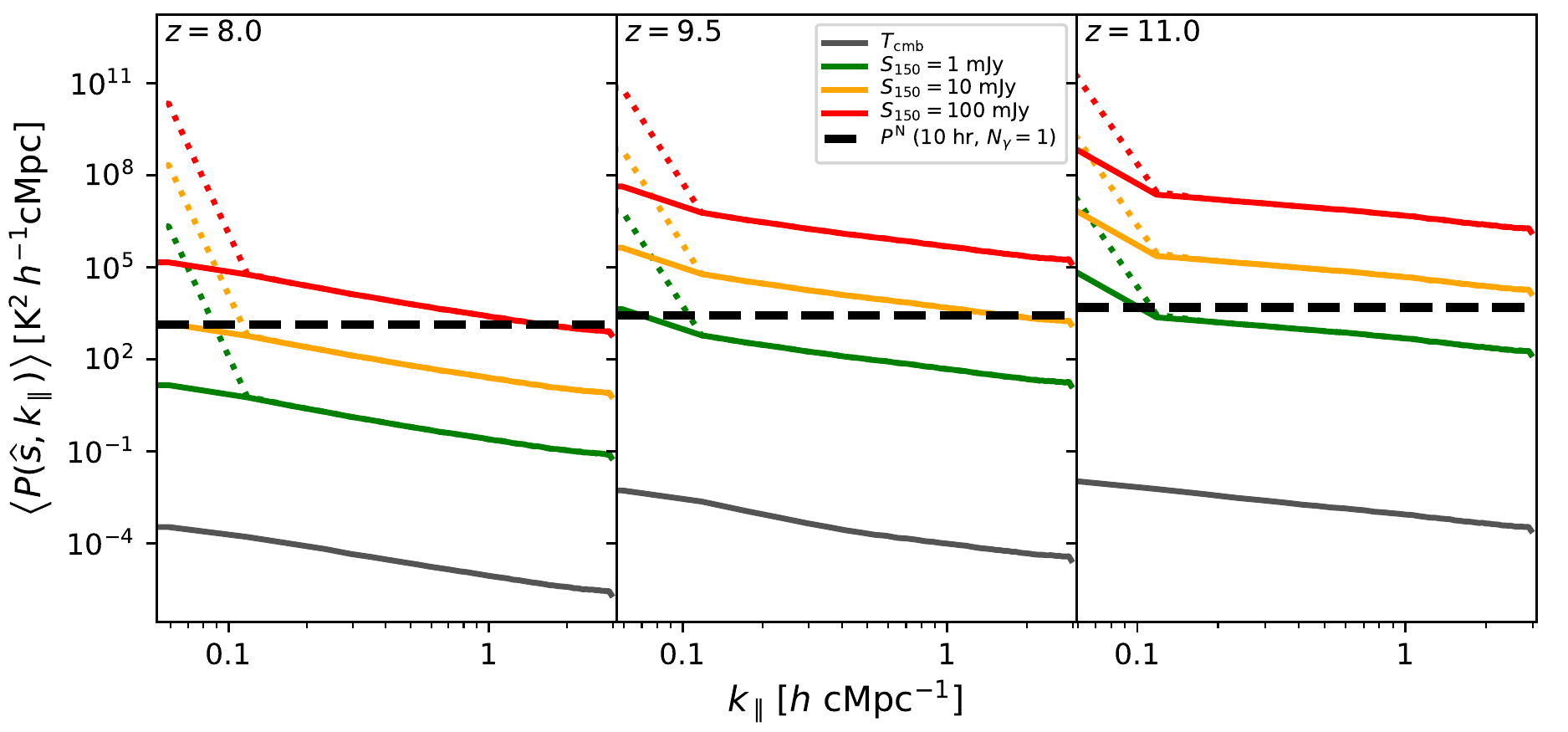}} \\
\subfloat[][EoS model 2 (\tt{FAINT GALAXIES})\label{fig:pspec_faint_gal}]{\includegraphics[width=\textwidth]{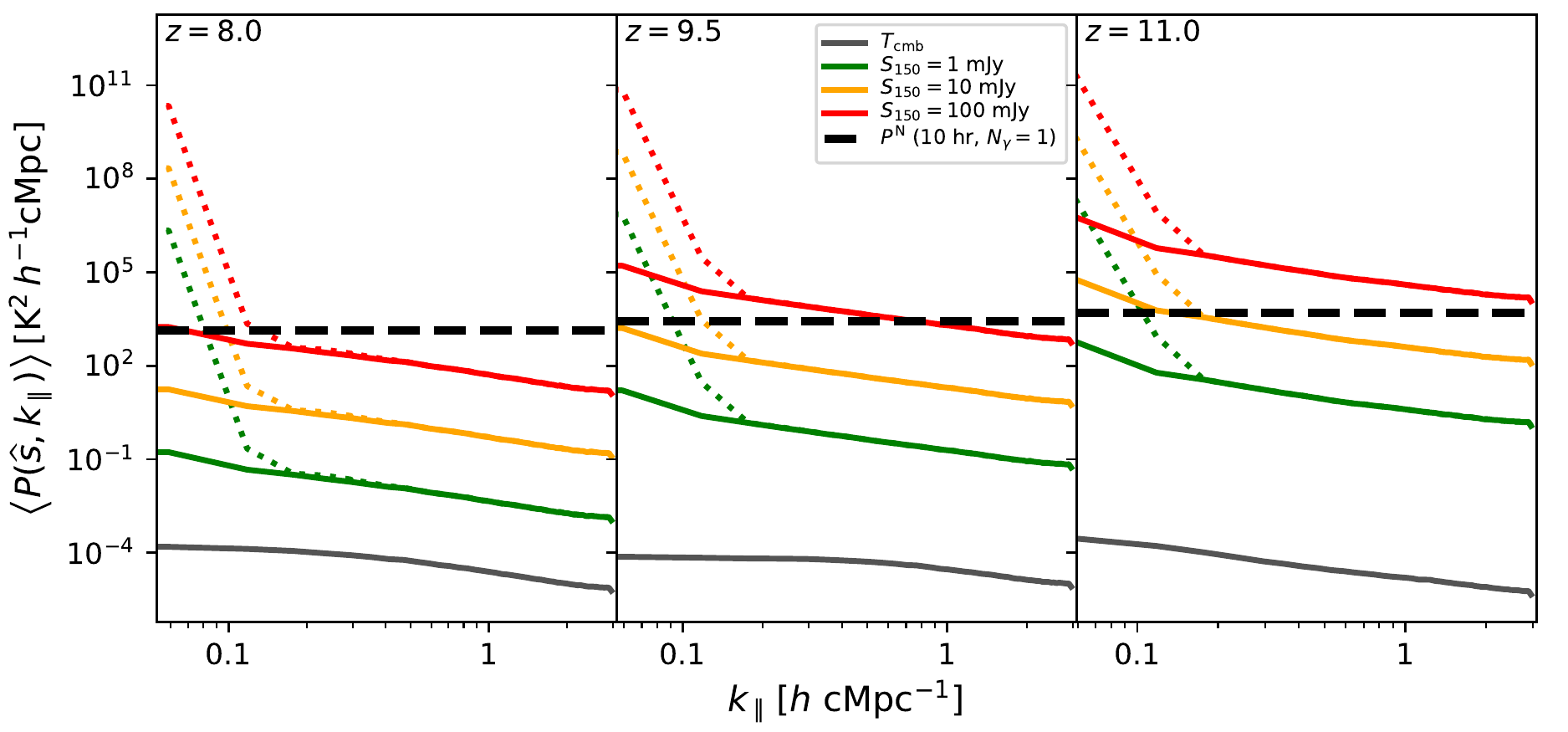}}
\caption{The expected value of the 1D line-of-sight power spectra of the differential flux densities in Figure~\ref{fig:dS_obs_redshifts} obtained by marginalizing Equation~(\ref{eqn:comoving-PS}) over $\hat{\boldsymbol{s}}$ in the three redshifted spectral subbands corresponding to $z=8$ (left), $z=9.5$ (middle), and $z=11$ (right). The black dashed line is the standard deviation of thermal uncertainty in the cross-power spectrum, $P^\textrm{N}$, obtained from Equation~(\ref{eqn:cross-power-noise-rms-avg}) corresponding to an integration time $\delta t=2\,\delta t_{0.5}=10$~hr along the sightline to a single source of compact background radiation ($N_\gamma=1$). The solid lines in green ($S_{150}=1$~mJy), orange ($S_{150}=10$~mJy), and red ($S_{150}=100$~mJy) correspond to when the continuum spectrum of the foreground and background radiation has been subtracted, while the dotted lines denote the corresponding 1D power spectra without such a continuum subtraction. The latter is shown to demonstrate the typical $k_\parallel$-modes that are contaminated when the continuum from foreground and background sources of radiation is not subtracted properly. For comparison, the 1D power spectra from fluctuations (absorption) against compact background radiation sources are stronger than those against the CMB (gray) by at least a few orders of magnitude. Consistent with Figure~\ref{fig:dS_obs_redshifts}, the 1D power spectra become stronger with increasing redshift. \label{fig:pspec_redshifts}}
\end{figure*}

Figure~\ref{fig:pspec_bright_gal} and Figure~\ref{fig:pspec_faint_gal} show the expected values of the 1D power spectra, $P(k_\parallel)\equiv\left\langle P(\hat{\boldsymbol{s}},k_\parallel)\right\rangle$, by marginalizing Equation~(\ref{eqn:comoving-PS}) over $\hat{\boldsymbol{s}}$, for the EoS models 1 and 2, respectively, for different values of $S_{150}$. The marginalization over $\hat{\boldsymbol{s}}$ is done to obtain the true expected value and should not be confused with the use of $N_\gamma$, which is used to obtain the best estimate of this expected value using as many independent lines ($N_\gamma$) of sight as available through actual observations. The latter will be investigated later. The gray curve denotes when $S_\gamma(\hat{\boldsymbol{s}},\nu,z=0) = S_\textrm{CMB}(\nu,z=0)$ in Equation~(\ref{eqn:S_rad}). The black dashed curve denotes the standard deviation of the bias-free noise cross-power, $P^\textrm{N}$, in Equation~(\ref{eqn:cross-power-noise-rms-avg}) using $\delta t=2\,\delta t_{0.5}=10$~hr and $N_\gamma=1$. Note that the motivation for using $N_\gamma=1$ here is just to compare 1D power spectra to direct detection and that the advantages of using multiple sightlines ($N_\gamma >1$) will be discussed later. The dotted and solid lines (green, orange, and red) show when $S_\textrm{obs}(\hat{\boldsymbol{s}},\nu)$ (with the continuum spectrum from the compact background source included) and the differential flux density, $\Delta S_\textrm{obs}(\hat{\boldsymbol{s}},\nu)$, are used to obtain the power spectrum, respectively, as detailed in \S\ref{sec:pspec}. It shows, for reference, the {\it Fourier} modes (in dotted lines) that could potentially be contaminated by the continuum radiation from both the foregrounds and the background along $\hat{\boldsymbol{s}}$ if not removed properly. 

In the EoS model 1, the 1D power spectrum at $z=8$ with an $S_{150}=100$~mJy compact background radiation source is detectable on large scales ($k_\parallel \lesssim 0.1\,h\,$cMpc$^{-1}$) with an S/N of $\sim 100$ and $k_\parallel \lesssim 0.5\,h$~cMpc$^{-1}$ with $\textrm{S/N} \gtrsim 5$. In contrast, in the direct detection scenario, even with complete coherent averaging over $\delta t=10$~hr, the absorption features in the spectra are detected with at best with an $\textrm{S/N} \lesssim 20$ and several broad structures remain undetectable. Absorption against fainter background sources ($S_{150}\le 10$~mJy) remains undetectable in both approaches. At $z=9.5$, the power spectrum with an $S_{150}=100$~mJy compact background radiation source is detectable on all scales with $\textrm{S/N} \gtrsim 100$ and at $k_\parallel \lesssim 0.5\,h\,$cMpc$^{-1}$ with $\textrm{S/N} \gtrsim 1000$. The corresponding scenario with $S_{150}=100$~mJy in a direct detection approach achieves an $\textrm{S/N} \lesssim 100$ at best while still being unable to detect some narrow features. Absorption against an $S_{150}=10$~mJy source is directly detectable with $\textrm{S/N} \sim 10$ on the largest scales but many of the narrow features are undetectable. With a 1D power spectrum, an $\textrm{S/N} \gtrsim 5$ is achieved for $k_\parallel\lesssim 0.6\,h\,$cMpc$^{-1}$ and $\textrm{S/N} \gtrsim 10$ for $k_\parallel\lesssim 0.3\,h\,$cMpc$^{-1}$. The $S_{150}=1$~mJy case is undetectable in both approaches. At $z=11$, direct detection of absorption features is possible in the $S_{150}=100$~mJy and 10~mJy cases with $\textrm{S/N} \sim 200$ and $\sim 20$, respectively, in most parts of the spectrum except on the narrowest scales. The $S_{150}=1$~mJy case is only marginally observable with $\textrm{S/N} \lesssim 3$ on the largest scales while none of the medium- and small-scale features is detectable. With a 1D power spectrum, for $S_{150}=100$~mJy and 10~mJy, all scales with $k_\parallel\lesssim 2\,h\,$~cMpc$^{-1}$ are detectable with $\textrm{S/N} \gtrsim 500$ and $\gtrsim 5$ respectively. Larger scales such as $k_\parallel\lesssim 1\,h\,$~cMpc$^{-1}$ are detectable with $\textrm{S/N} \gtrsim 2000$ and $\gtrsim 20$, respectively. With $S_{150}=1$~mJy, the 1D power spectrum is detectable with $\textrm{S/N} \gtrsim 5$ at $k_\parallel\lesssim 0.08\,h\,$cMpc$^{-1}$. 

In contrast, the EoS model 2 is harder to detect in general. At $z=8$, neither approach is able to yield a detection for any $S_{150}$ considered here. At $z=9.5$, some medium to large scales with $S_{150}=100$~mJy are detectable in a direct detection approach with $\textrm{S/N} \sim 5$--10. In 1D power spectrum, spatial scales with $k_\parallel\lesssim 0.2\,h\,$cMpc$^{-1}$ and $k_\parallel\lesssim 0.1\,h\,$cMpc$^{-1}$ are detected with $\textrm{S/N} \gtrsim 5$ and $\gtrsim 10$, respectively. Absorption against fainter values, $S_{150}\le 10$~mJy, is undetectable in both approaches at this redshift. At $z=11$, direct detection of medium- to large-scale features is possible with $\textrm{S/N} \sim 10$--100 in the $S_{150}=100$~mJy case, whereas none of the small-scale features is detected. Some of the broadest features are detected against the  $S_{150}=10$~mJy background source with $\textrm{S/N} \lesssim 5$ while none of the features is detected for the $S_{150}=1$~mJy case. With the 1D power spectrum, all scales with $k_\parallel\lesssim 2\,h\,$cMpc$^{-1}$, $k_\parallel\lesssim 0.8\,h\,$cMpc$^{-1}$, and $k_\parallel\lesssim 0.2\,h\,$cMpc$^{-1}$ are detected with $\textrm{S/N} \gtrsim 5$, $\gtrsim 10$, and $\gtrsim 100$, respectively, in the $S_{150}=100$~mJy case. Even for the $S_{150}=10$~mJy case, the 1D power spectrum on $k_\parallel\lesssim 0.08\,h\,$cMpc$^{-1}$ is detected with $\textrm{S/N} \gtrsim 5$. The 1D power spectrum on all scales in the $S_{150}=1$~mJy case remains undetectable. 

In summary, both the EoS models exhibit an intrinsic increase in the strength of the 1D power spectrum with redshift where the absorption features appear stronger. This is due to an increase in both the optical depth (from decreasing spin temperature) and the background radiation (due to the spectral index) at lower frequencies (higher redshifts). When compared to a direct detection approach, the 1D power spectrum offers much improved sensitivity and even makes detection on previously inaccessible scales possible. The detectability of the 1D power spectrum can be potentially improved further by a factor $\simeq\sqrt{N_\gamma}$ by averaging over as many sightlines as available toward known compact sources of background radiation such as AGNs, star-forming radio galaxies, and radio afterglows from GRBs, which will reduce both the cosmic variance and the thermal noise power. This especially improves sensitivity to the smallest scales (large $k_\parallel$-modes). If nominal variations in $N_\textrm{a}A_\textrm{e}/T_\textrm{sys}$ of 20\% lower (higher) values at the lowest (highest) frequency subbands are considered, the dependence on array sensitivity as $\sim (N_\textrm{a}A_\textrm{e}/T_\textrm{sys})^{-2}$ will result in a correction factor of 1.5625 ($\approx 0.7$) to $P^\textrm{N}$ relative to the nominal values.

It must be noted that proper removal or suppression of contamination, especially on large scales (small $k_\parallel$-modes), caused by the continuum both from the foreground and the background sources of radiation, is a prerequisite for the success of not only the 1D power spectrum but also the direct detection approach as they could cause power leakage and contamination onto other scales (larger $k_\parallel$-modes). The sidelobes from the synthesized PSF will also contain significant spectral structures that will contaminate a range of $k_\parallel$-modes. The effects of such a contamination and constraints on the quality of the synthesized PSF are discussed later. Spectral gaps due to RFI flagging or unflagged RFI, both of which abound at low radio frequencies, will also leak spurious power into the various $k_\parallel$-modes. Careful strategies for RFI flagging and removal \citep[for example,][]{par12b,bha19} are needed to mitigate such effects.

In practice, there will be a distribution of flux densities of background radiation sources at any redshift. Thus, observations toward each of the compact background radiation sources will appear with different S/N depending on the strengths of the background objects as illustrated by Figure~\ref{fig:dS_obs_redshifts}. A simple averaging of the 1D power spectra of the differential flux densities will not only make the interpretation complicated, but will also not result in an optimal S/N. Because this paper does not use a distribution but only constant values for flux densities for the background radiation sources, the issue of mixing different sensitivities does not arise. A simple outline of a 1D power spectrum approach using optical depths instead of flux densities is presented in \S\ref{sec:optical-depth-PS} to address this issue. Devising a detailed scheme is beyond the scope of this paper and will be explored in future work. The intent of this paper is to present a guiding framework for a statistical approach based on the 1D power spectrum of redshifted 21~cm absorption by H~{\sc i} in the IGM at various redshifts by observing a compact background radiation source of a given average strength. 

\subsection{Higher-order Moments} \label{sec:pdf}

\begin{figure*}
\includegraphics[width=\linewidth]{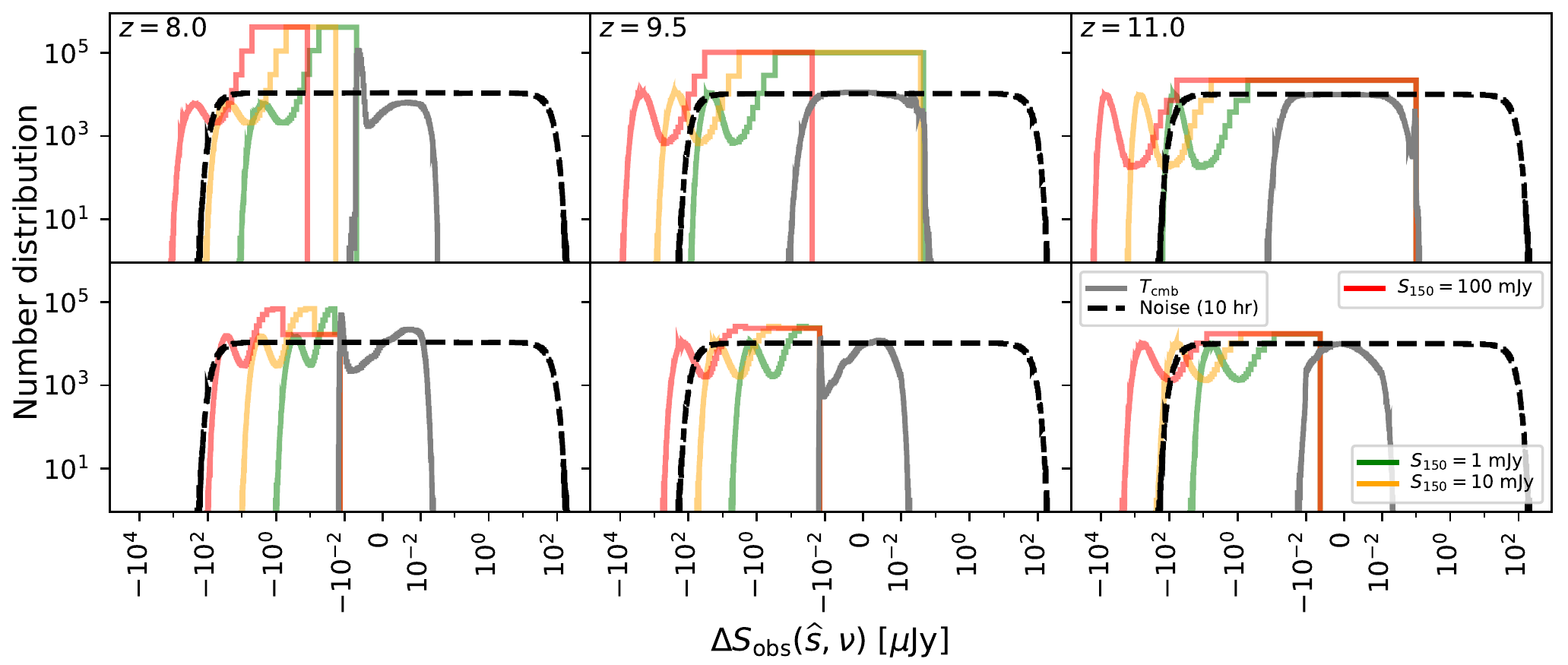}
\caption{{\it Top}: Number distribution of the differential flux densities in the spectra along sightlines that contain a compact source of background radiation (green, orange, and red), the CMB only (gray), and thermal noise fluctuations for an integration time of $\delta t=10$~hr (black dashed). The fluctuations in the case of compact background radiation sources are mostly seen in absorption unlike that against the CMB. Except for the thermal noise fluctuations which follow a Gaussian distribution of zero mean, the signal fluctuations are not Gaussian and are heavily skewed. The fraction of voxels that lie outside the thermal noise envelope denote those that can be directly detected in a 10~hr observation along a single sightline. The skewed distributions demonstrate that the higher-order moments contain important information about non-Gaussian statistics, which cannot be captured by variance or the 1D power spectrum alone. The $x$-axis is logarithmically scaled along both positive and negative directions. {\it Bottom}: Same as the top panels but for the {\tt FAINT GALAXIES} model (EoS model 2). 
\label{fig:fluxdensity_histogram}}
\end{figure*}

Alternate statistical metrics besides the power spectrum such as skewness and kurtosis could be used to detect non-Gaussian features in the spectral structures by examining the distribution of flux densities in the absorption spectra for higher-order moments \citep{wat14,wat15,kit18a,kit18b}. Figure~\ref{fig:fluxdensity_histogram} shows the histograms of the flux density spectra for the EoS models 1 (top) and 2 (bottom) at the three redshifts ($z$ increases toward right) for different values of $S_{150}$ characterizing the compact background radiation source model as well as the CMB. The histogram for thermal noise from $\delta t=10$~hr integration, which is a Gaussian, is also shown. The same color coding and line style apply as in previous figures. The fraction of voxels that lie outside the envelope of the noise histogram can be detected directly in the spectrum. 

The flatness and wide bins in some portions of the histograms result from the adaptive binning algorithm used. Regardless, it is clearly noted that each of these distributions is distinctly different from one another, which provides a handle to detect the higher-order moments and distinguish between the underlying models. This paper does not focus on the examination of these higher-order moments and is left to future work. 

\section{Requirement on Number Count of High-redshift Background Sources}\label{sec:N_gamma_required}

The discussion in \S\ref{sec:pspec} mostly used $N_\gamma=1$ and a nominal $\delta t=10$~hr. Here, the observational implications for the minimum number count of the compact background radiation sources at high redshifts is examined with the requirement that all the $k_\parallel$-modes accessible through the 1D power spectrum are detectable with a minimum S/N detection threshold, $\rho$. In other words, the minimum number of compact background sources that will be required for a high-significance detection of the 1D power spectrum in all accessible $k_\parallel$-modes will be determined. In this paper, the nominal detection threshold chosen is $\rho=5$. 

\begin{figure*}
\centering
% \subfloat[][EoS model 1 (\tt{BRIGHT GALAXIES})\label{fig:N_gamma_bright_gal}]{\includegraphics[width=\textwidth]{figures/Nqso_required_Bright_galaxies_fiducial_1024.pdf}} \\
% \subfloat[][EoS model 2 (\tt{FAINT GALAXIES})\label{fig:N_gamma_faint_gal}]{\includegraphics[width=\textwidth]{figures/Nqso_required_Faint_galaxies_fiducial_1024.pdf}}
\subfloat[][EoS model 1 (\tt{BRIGHT GALAXIES})\label{fig:N_gamma_bright_gal}]{\includegraphics[width=\textwidth]{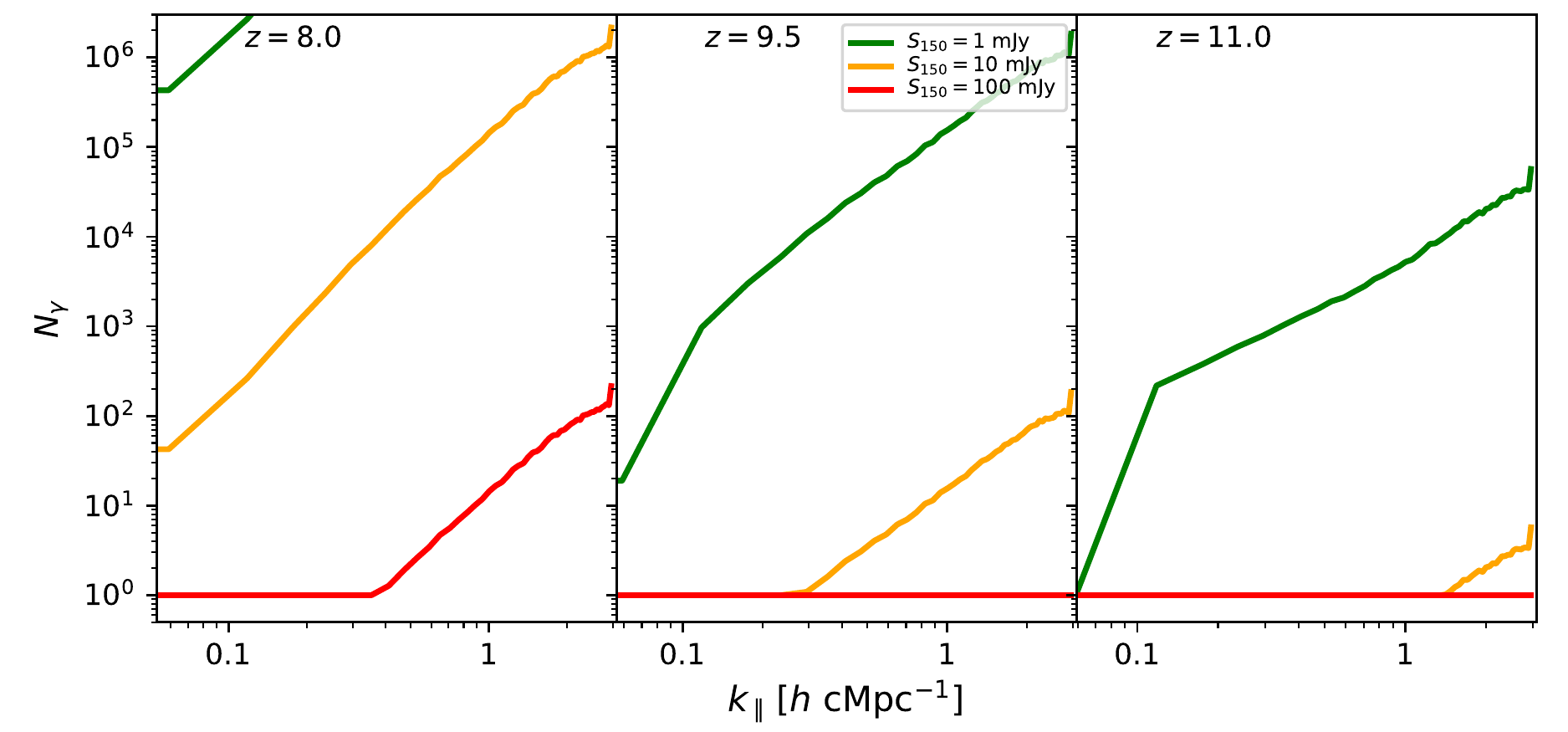}} \\
\subfloat[][EoS model 2 (\tt{FAINT GALAXIES})\label{fig:N_gamma_faint_gal}]{\includegraphics[width=\textwidth]{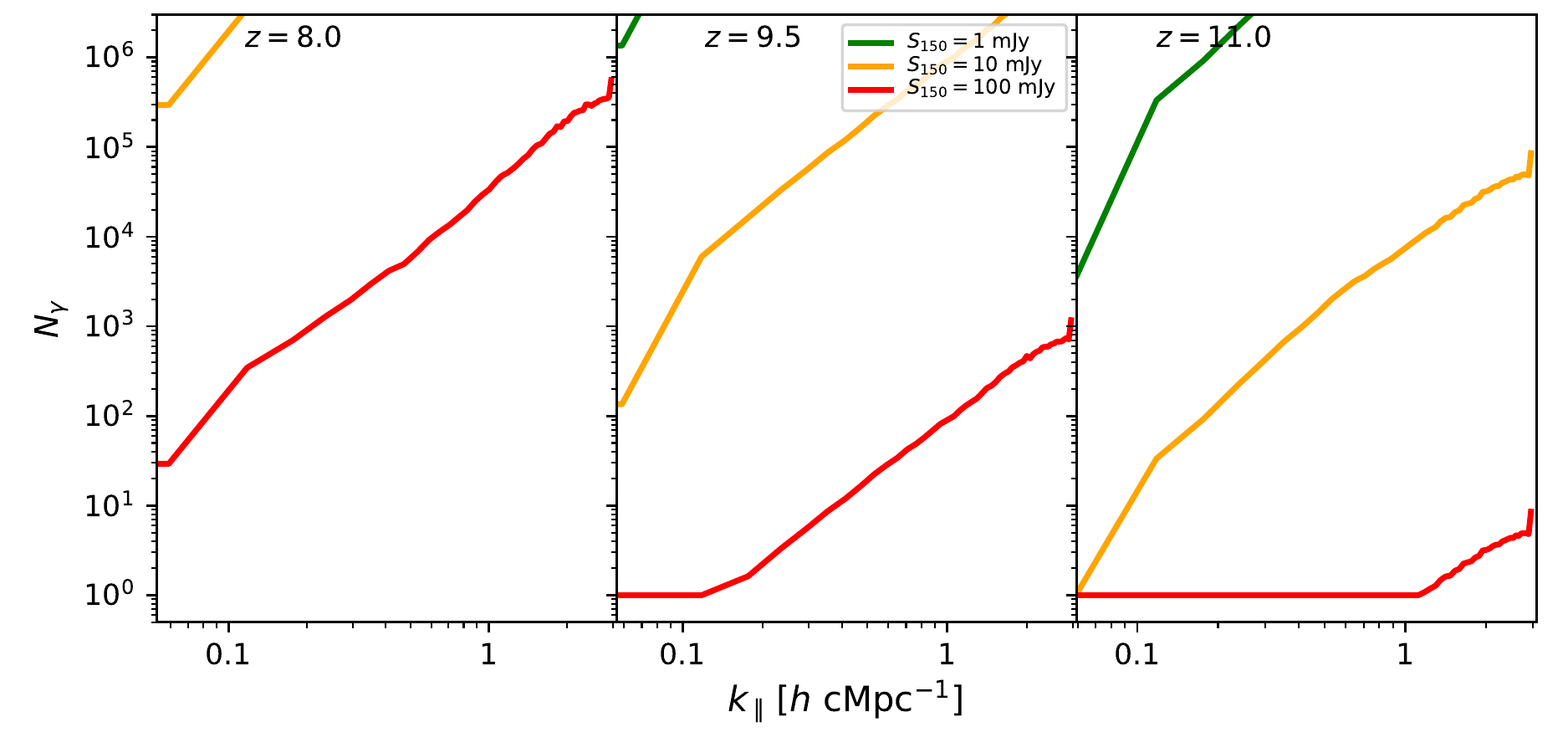}}
\caption{The minimum number count of compact background radiation sources, $N_\gamma$, required at redshifts $>z$ for detecting the 1D line-of-sight power spectrum with $\textrm{S/N} \ge \rho$ in each line-of-sight spatial scale (or $k_\parallel$-mode) based on Equation~(\ref{eqn:N_gamma}) at various redshifts and flux densities $S_{150}$ using an array sensitivity of $\eta N_\textrm{a}A_\textrm{e}/T_\textrm{sys}=800\,\textrm{m}^2\,\textrm{K}^{-1}$ (anticipated for SKA1-low) and nominal values of other parameters in Table~\ref{tab:nominal-values}. $N_\gamma=1$ is the minimum number of compact background radiation sources required for this study. Because the power in the fluctuations become stronger with increasing redshift while $P^N$ only increases weakly, the $N_\gamma$ required decreases with increasing redshift. \label{fig:N_gamma}}
\end{figure*}

By requiring that $\langle P(\hat{\boldsymbol{s}},k_\parallel)\rangle / P^\textrm{N} \ge \rho$ at all accessible $k_\parallel$-modes, Equation~(\ref{eqn:cross-power-noise-rms-avg}) yields: 
\begin{align}\label{eqn:N_gamma}
    N_\gamma &\ge \left[\frac{\rho}{\langle P(\hat{\boldsymbol{s}},k_\parallel)\rangle}\,\frac{\Delta r_z}{2\,\Delta\nu_z\,\delta t_{0.5}}\right]^2 \left(\frac{\lambda_z^2}{\Omega}\,\frac{T_\textrm{sys}}{\eta N_\textrm{a}A_\textrm{e}}\right)^4.
\end{align}
This establishes the scaling relations to obtain the minimum number of compact background sources of radiation required at $z_\gamma>z$ for detecting the 1D power spectrum at redshift $z$ with a given significance threshold, $\rho$. Because at least one object is required to be observed, $N_\gamma\ge 1$. Whenever $N_\gamma=1$, it implies that just one observation of the spectrum against the compact background radiation source with integration time $\le \delta t$ is sufficient to obtain a power spectrum with $\textrm{S/N}\geq\rho$. An approximately equivalent inference is that direct detection of the absorption features with integration time $\le \delta t$ is plausible on the line-of-sight spatial scales corresponding to those $k_\parallel$-modes where $N_\gamma=1$. Figure~\ref{fig:N_gamma_bright_gal} and \ref{fig:N_gamma_faint_gal} show the minimum $N_\gamma$ required at various redshifts for detecting the 1D power spectrum in all accessible $k_\parallel$-modes in the EoS models 1 and 2 respectively using the nominal values for various parameters (see Table~\ref{tab:nominal-values}).  

Because of intrinsically higher power, the requirement on $N_\gamma$ is less severe in the EoS model 1. With increasing redshift, statistical detection of absorption features even against weaker background sources becomes plausible with lower $N_\gamma$ because of the inherent increase in the 1D power spectrum strength at higher redshifts. For example, with $N_\gamma=100$ at $z=9.5$ and $z=11$, power spectra for even the $S_{150}=1$~mJy case become detectable at $k_\parallel\lesssim 0.08\,h\,$cMpc$^{-1}$ and $k_\parallel\lesssim 0.1\,h\,$cMpc$^{-1}$, respectively, and the $S_{150}=10$~mJy case is detectable at all $k_\parallel$-modes with $\textrm{S/N} \gtrsim \rho$. The $S_{150}=100$~mJy case is detectable at all redshifts. In EoS model 2, the same qualitative trends hold but the overall requirement on $N_\gamma$ is more severe. For example, even with $N_\gamma=100$, the $S_{150}=100$~mJy case is only detectable at $k_\parallel\lesssim 0.08\,h\,$cMpc$^{-1}$ at $z=8$. At $z=9.5$, the power on $k_\parallel\lesssim 1\,h\,$cMpc$^{-1}$ is detectable for $S_{150}=100$~mJy, whereas the $S_{150}=10$~mJy case is not detectable on any scales. At $z=11$, with $N_\gamma=100$, the $S_{150}=10$~mJy case is detectable up to $k_\parallel\lesssim 0.2\,h\,$cMpc$^{-1}$, whereas the $S_{150}=1$~mJy case is not detectable on any scale. 

Note that the total observing time is given by $\Delta t \le N_\gamma\,\delta t$. For nominal values of $N_\gamma=100$ and $\delta t=10$~hr, the total observing time is nominally $\Delta t\le 1000$~hr. The inequality applies when more than one background radiation source lies in the same field of view. 

Here, the minimum required $N_\gamma$ is determined without any {\it a priori} knowledge of the number density and evolution of the population of compact background sources at high redshifts. In the case of AGNs, the presence of a significant population of high redshifts at low radio frequencies is yet to be confirmed observationally. The dearth of bright radio AGNs could potentially arise due to either a bias against radio signatures in those selected optically, or significant inverse-Compton (IC) losses against the brighter CMB at these high redshifts. Radio-based criteria that are more efficient at selecting such high-redshift objects with radio signatures are being explored \citep{sax18}. Background AGNs faint in radio frequencies with $S_{150}\lesssim 10$~mJy are expected to be quite abundant (possibly hundreds to thousands) at high redshifts \citep{hai04} relative to brighter ones even after accounting for the IC losses \citep{sax17}; these will become accessible with surveys with the EGMRT, LOFAR, and the SKA. For example, \citet{hai04} predict $\sim 2000$ quasars at flux densities $\sim 6$~mJy at $\sim 100$~MHz available over the full sky at $8<z<12$ and the Very Large Array Faint Images of the Radio Sky at Twenty cm \citep[VLA FIRST survey;][]{hel15} may have already detected $\sim 10^3$--$10^4$ quasars at $\sim 1$~mJy flux densities at 1.4~GHz at $z\gtrsim 7$. Thus, $N_\gamma\gtrsim 100$ AGNs with $S_{150}\lesssim 10$~mJy beyond each of the redshifts analyzed here appear to be plausible based on these models.

It is noted that $N_\gamma$ depends sensitively on the array sensitivity as $\sim (N_\textrm{a}A_\textrm{e}/T_\textrm{sys})^{-4}$. If a systematic variation of $\approx 20$\% lower (higher) values relative to the nominal value is considered in the array sensitivity at the lowest (highest) spectral subbands, the minimum number of background sources required gets significantly modified to $N_\gamma^\prime\approx2.44\,N_\gamma$ ($\approx 0.48\,N_\gamma$). This implies that the detection should be possible with a correspondingly higher (lower) number of background sources at the lowest (highest) spectral subbands compared to the nominal estimates, thus making it a stricter and a more lenient lower limit on $N_\gamma$ at the lowest and highest spectral subbands, respectively. However, this effect is compensated by the increase in the intrinsic strength of the 1D power spectrum with redshift. 

\section{Requirement on array sensitivity}\label{sec:array-sensitivity-required}

\begin{figure*}
\centering
% \subfloat[][EoS model 1 (\tt{BRIGHT GALAXIES})\label{fig:Aeff_over_T_bright_gal}]{\includegraphics[width=\textwidth]{figures/A_over_T_required_Bright_galaxies_fiducial_1024.pdf}} \\
% \subfloat[][EoS model 2 (\tt{FAINT GALAXIES})\label{fig:Aeff_over_T_faint_gal}]{\includegraphics[width=\textwidth]{figures/A_over_T_required_Faint_galaxies_fiducial_1024.pdf}}
\subfloat[][EoS model 1 (\tt{BRIGHT GALAXIES})\label{fig:Aeff_over_T_bright_gal}]{\includegraphics[width=\textwidth]{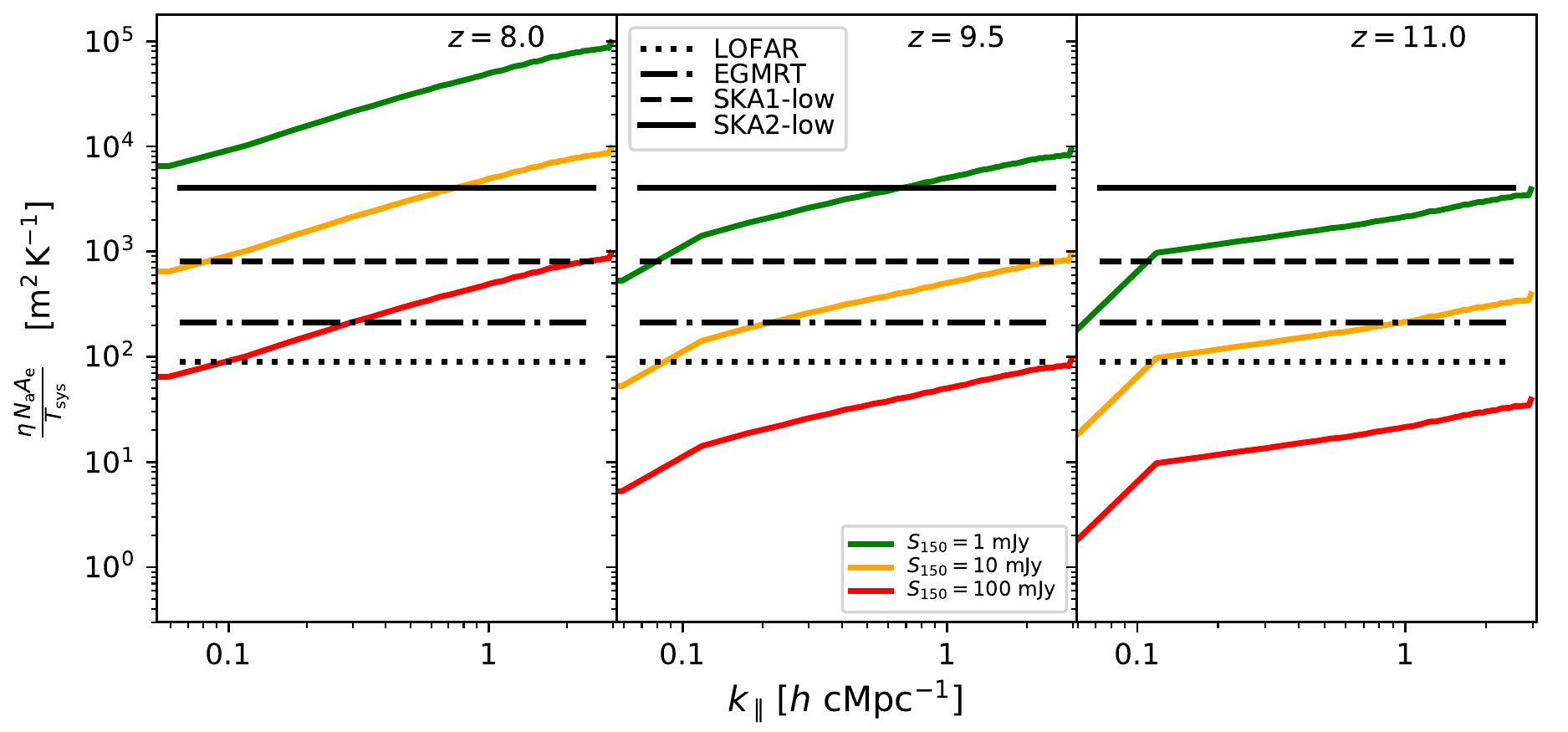}} \\
\subfloat[][EoS model 2 (\tt{FAINT GALAXIES})\label{fig:Aeff_over_T_faint_gal}]{\includegraphics[width=\textwidth]{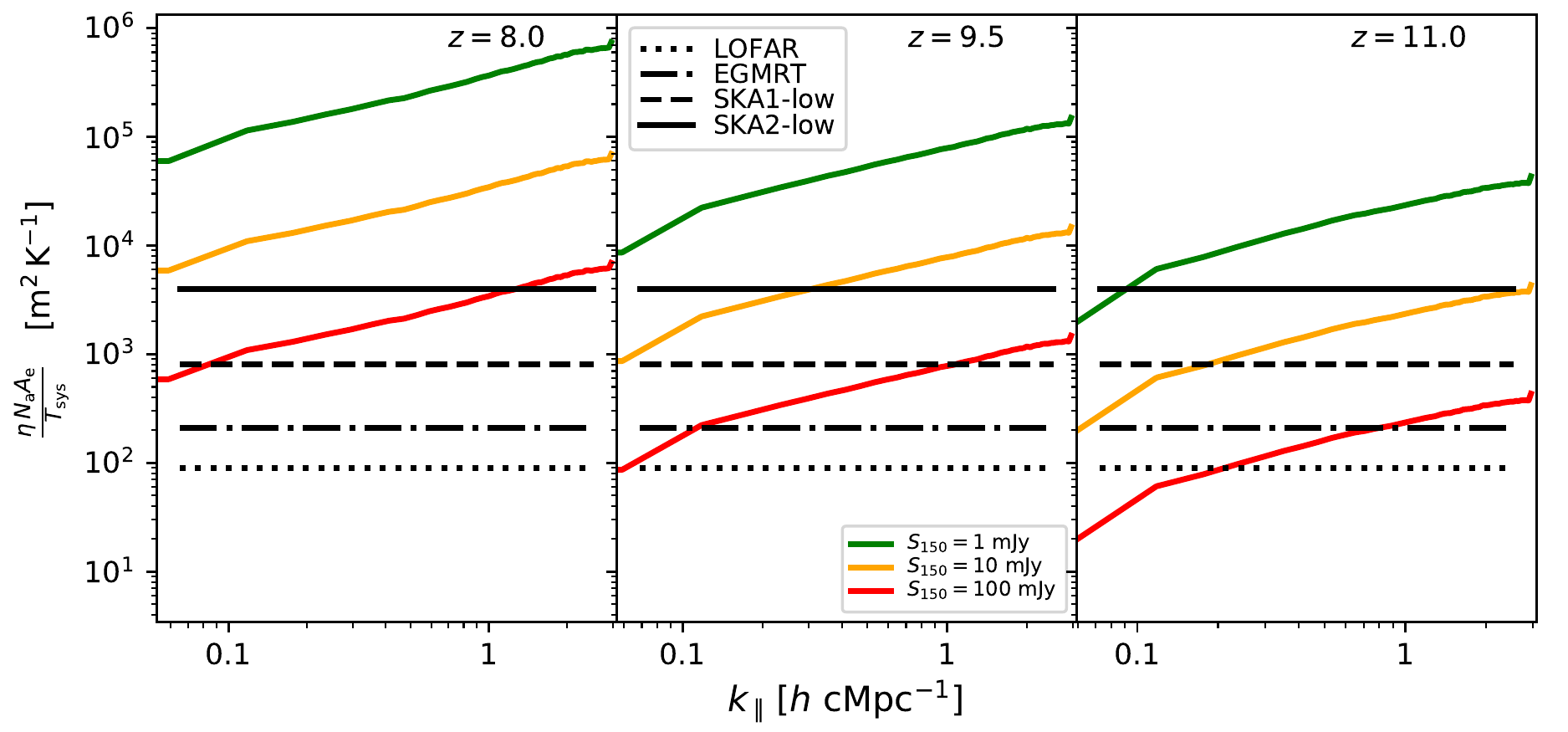}}
\caption{The minimum array sensitivity, $\eta N_\textrm{a}A_\textrm{e}/T_\textrm{sys}$, from Equation~(\ref{eqn:Aeff_over_T}) required for detecting the 1D power spectrum in each of the $k_\parallel$-modes with $\textrm{S/N} \ge \rho$ at various redshifts using an integration time of $\delta t=10$~hr each on $N_\gamma=100$ compact background radiation sources, shown for different background radiation strengths and nominal values of other parameters listed in Table~\ref{tab:nominal-values}. The current or anticipated array sensitivity performance of some current and planned interferometer arrays are shown for reference using the black lines. While even current instruments such as LOFAR with relatively the lowest sensitivity are significantly capable of detecting the 1D power spectrum against strong and even moderately weak sources of background radiation with high significance on certain $k$-modes and selected redshifts, the next generation instruments like the SKA will enhance this capability by at least an order of magnitude and thus have the capability of detecting absorption even against weak background objects with $S_{150}\simeq 1$~mJy. Due to inherently increasing power at higher redshifts, the sensitivity requirement becomes correspondingly less severe in those spectral subbands. \label{fig:Aeff_over_T}}
\end{figure*}

A requirement on instrument performance can be placed if the combination of parameters, $N_\gamma$ and $\delta t=\Delta t/N_\gamma$ (assuming each field of view contains only one background source), is specified. By rearranging Equation~(\ref{eqn:N_gamma}), the minimum array sensitivity required is:
\begin{align}\label{eqn:Aeff_over_T}
    \frac{\eta N_\textrm{a}A_\textrm{e}}{T_\textrm{sys}} &\ge \frac{\lambda_z^2/\Omega}{N_\gamma^{1/4}}\left[\frac{\rho}{\langle P(\hat{\boldsymbol{s}},k_\parallel)\rangle}\,\frac{\Delta r_z}{2\,\Delta\nu_z\,\delta t_{0.5}}\right]^{1/2}.
\end{align}
For the nominal parameter values listed in Table~\ref{tab:nominal-values}, assuming the number of observations of compact background radiation sources is $N_\gamma=100$, Figure~\ref{fig:Aeff_over_T_bright_gal} and Figure~\ref{fig:Aeff_over_T_faint_gal} show the required interferometer array sensitivity including the system efficiency parameterized by $\eta N_\textrm{a}A_\textrm{e}/T_\textrm{sys}$ in order to detect the power spectrum in the EoS models 1 and 2, respectively in all available $k_\parallel$-modes with $\textrm{S/N} \ge \rho$ for different redshifts and background radiation strengths, $S_{150}$. Also shown are the anticipated array sensitivity values for the LOFAR, the proposed EGMRT, the upcoming SKA1-low, and the eventual SKA2 telescopes. 

For the nominal values chosen, regions of the plots below the telescope sensitivity parameter are to be interpreted as detectable with $\textrm{S/N} \ge \rho$. The nominal observing time per target is 10~hr, and the total observing time is 1000~hr for $N_\gamma=100$ targets. In both the EoS models, a higher sensitivity is required to detect the small-scale structures (higher $k_\parallel$-modes) because of the inherently smaller power in those scales relative to the larger scales. The sensitivity requirement is less severe at higher redshifts because the intrinsic strength of the 1D power spectrum of the H~{\sc i} absorption features increases significantly with redshift.

LOFAR will be able to detect the 1D power spectrum of the EoS model 1 for $S_{150}=100$~mJy at $k_\parallel\lesssim 0.1\,h\,$cMpc$^{-1}$ at $z=8$, and all scales at $z=9.5$ and $z=11$. At $z=9.5$ and $z=11$, LOFAR can detect the power spectrum at $k_\parallel\lesssim 0.08\,h\,$cMpc$^{-1}$ and $k_\parallel\lesssim 0.1\,h\,$cMpc$^{-1}$ respectively, for $S_{150}=10$~mJy. The EGMRT can detect the power spectrum for $S_{150}=100$~mJy at $k_\parallel\lesssim 0.3\,h\,$cMpc$^{-1}$ at $z=8$, and all scales at $z=9.5$ and $z=11$. It can detect power spectra for $S_{150}=10$~mJy on scales $k_\parallel\lesssim 0.2\,h\,$cMpc$^{-1}$ and $k_\parallel\lesssim 1\,h\,$cMpc$^{-1}$ at $z=9.5$ and $z=11$, respectively. SKA1-low can detect almost all the scales in the power spectrum at all redshifts for $S_{150}=100$~mJy. It will detect the $S_{150}=10$~mJy case on $k_\parallel\lesssim 0.08\,h\,$cMpc$^{-1}$ at $z=8$ and on almost all the scales at $z=9.5$ and $z=11$. It will even detect the $S_{150}=1$~mJy case on $k_\parallel\lesssim 0.08\,h\,$cMpc$^{-1}$ and $k_\parallel\lesssim 0.1\,h\,$cMpc$^{-1}$ at $z=9.5$ and $z=11$, respectively. The eventual SKA2 will detect power on all scales at all three redshifts for the $S_{150}=100$~mJy case, while for the $S_{150}=10$~mJy case it will be able to detect on $k_\parallel\lesssim 0.7\,h\,$cMpc$^{-1}$ at $z=8$ and on all $k_\parallel$-modes at $z=9.5$ and $z=11$. The $S_{150}=1$~mJy case will be detectable on $k_\parallel\lesssim 0.7\,h\,$cMpc$^{-1}$ at $z=9.5$ and on all $k_\parallel$-modes at $z=11$ with the SKA2. 

LOFAR will be able to detect power in the EoS model 2 for $S_{150}=100$~mJy only at $z=11$ on $k_\parallel\lesssim 0.2\,h\,$cMpc$^{-1}$, whereas the EGMRT can detect power on $k_\parallel\lesssim 0.1\,h\,$cMpc$^{-1}$ and $k_\parallel\lesssim 0.8\,h\,$cMpc$^{-1}$ at $z=9.5$ and $z=11$, respectively. The SKA1-low will improve detectability of the $S_{150}=100$~mJy case at $z=8$, $z=9.5$, and $z=11$ to $k_\parallel\lesssim 0.08\,h\,$cMpc$^{-1}$, $k_\parallel\lesssim 1\,h\,$cMpc$^{-1}$, and all scales, respectively. The SKA2 will improve this further to $k_\parallel\lesssim 1\,h\,$cMpc$^{-1}$ at $z=8$, and all scales at $z=9.5$ and $z=11$. Further, it will not only be able to detect the $S_{150}=10$~mJy case on $k_\parallel\lesssim 0.3\,h\,$cMpc$^{-1}$ and all scales at $z=9.5$ and $z=11$, respectively, but also the $S_{150}=1$~mJy case on $k_\parallel\lesssim 0.09\,h\,$cMpc$^{-1}$ at $z=11$.

In summary, despite some scales in the reionization models being difficult to detect in the 1D power spectrum, most of the current and planned interferometer arrays have promising prospects of detecting line-of-sight 1D power spectrum on various spatial scales due to the improved sensitivity from a power spectrum approach, particularly on the smaller scales, that would otherwise not be possible in a direct detection approach alone. Note that if, coincidentally, some fields of view in an observation of duration $\delta t$ contain more than one compact background radiation source, the total observing time, $\Delta t$, and also the array sensitivity requirement can be correspondingly lowered. Thus, the estimates of array sensitivity here are conservative.  

\section{Effects of Chromatic PSF}\label{sec:sidelobes}

One of the systematic limitations in power spectrum approaches for detecting the redshifted 21~cm from the {\it cosmic dawn} and the EoR is the contamination from sidelobes in the synthesized PSF caused by gaps in the synthesized aperture and the confusing radio sources in the image encapsulated by $S_\textrm{fg}(\hat{\boldsymbol{s}},\nu,z=0)$ in Equation~(\ref{eqn:S_obs}). The in-voxel contribution to $S_\textrm{fg}(\hat{\boldsymbol{s}},\nu,z=0)$ due to the confusing sources at $\hat{\boldsymbol{s}}$ usually have smooth spectra and can be relatively easily isolated in the {\it Fourier} domain at small $k_\parallel$-modes (see dotted lines in Figure~\ref{fig:pspec_redshifts}). However, the sidelobe contributions from confusing sources in the entire field of view at $\hat{\boldsymbol{s}}$ have more spectral structure that makes them harder to isolate. It is referred to as {\it mode-mixing}, wherein the transverse structures in the sidelobes of the synthesized PSF vary as a function of frequency and thus manifest as spectral structures contaminating the $k_\parallel$-modes. The largest $k_\parallel$-mode so affected depends linearly on the largest $k_\perp$-mode in the measurements, and this linear dependence between the affected $k_\parallel$-modes and the $k_\perp$-modes is referred to as the {\it foreground wedge} \citep[for details, see][]{bow09,liu09,liu14a,liu14b,dat10,liu11,gho12,mor12,mor19,par12b,tro12,ved12,dil13,pob13,dil14,thy13,thy15a,thy15b,thy16}.

The equation describing the envelope of the $k_\parallel$-modes affected by {\it mode-mixing}, also known as the {\it foreground wedge}, is given by \citep{thy13}:
\begin{align}\label{eqn:wedge}
    k_{\parallel,\textrm{w}} &\simeq \frac{H_0 E(z)\,r_z}{c(1+z)}\,\sin{\left(\theta_\textrm{p}/2\right)}\,k_\perp,
\end{align}
where $\theta_\textrm{p}$ is the angular FWHM of the primary beam of the antenna power pattern. Using $k_\perp = 2\pi |u|/r_z$ \citep{mor04} and $1/|u|\simeq 2\sin{(\theta_\textrm{s}/2)}$,
\begin{align}
    k_{\parallel,\textrm{w}} &\simeq \frac{2\pi H_0 E(z)}{c(1+z)}\,\frac{\sin{(\frac{\theta_\textrm{p}}{2})}}{2\sin{(\theta_\textrm{s}/2)}} \\
                &\simeq \frac{2\pi H_0 E(z)}{c(1+z)}\,\frac{\sin{(\frac{\theta_\textrm{p}}{2})}}{\theta_\textrm{s}},
\end{align}
where the last equation resulted from using the small-angle approximation valid for $\theta_\textrm{s}\ll 1$. Using nominal values of $\theta_\textrm{p}=5\arcdeg$ and $\theta_\textrm{s}=10\arcsec$, $k_{\parallel,\textrm{w}}\approx 3.1\,h\,$cMpc$^{-1}$, $\approx 3.4\,h\,$cMpc$^{-1}$, and $\approx 3.6\,h\,$cMpc$^{-1}$ at $z=8$, 9.5, and 11, respectively. Thus, all the modes in this analysis will be contaminated by sidelobe confusion. Note that this is a result of coarse values chosen for $\delta\nu_z$, limited by $\delta r_z$ in the 21cmFAST simulations. If finer frequency channel widths are available, uncontaminated $k_\parallel$-modes beyond the {\it foreground wedge}, called the {\it EoR window}, will be available. In practice, a finer $\delta\nu_z$ can be chosen to make larger values of uncontaminated $k_\parallel$-modes available even in this 1D power spectrum approach. The choice of $\delta\nu_z$ and the resulting extent of sidelobe contamination into $k_\parallel$-modes applies to any approach, including direct detection.

A simple formalism is presented here to express the power spectrum of the sidelobes from the confusing sources in the synthesized image cube. The classical confusion from radio sources present in each pixel and the sidelobe response from all such residual confusing sources produce the sidelobe confusion at any given location and along the spectral axis. Assuming all sources brighter than five times the {\it rms} of confusion noise have been perfectly removed, the {\it rms} flux density in the residuals after integrating over the solid angle of the synthesized PSF is given by \citep{con12}:
\begin{align}\label{eqn:source-confusion-rms}
    \sigma_\textrm{c}\equiv\sigma_\textrm{c}(\nu_z,\theta_\textrm{s}) &\approx 1.2\,\mu\textrm{Jy} \left(\frac{\nu_z}{3.02\,\textrm{GHz}}\right)^{-0.7}\left(\frac{\theta_\textrm{s}}{8\arcsec}\right)^\frac{10}{3}
\end{align}
where it has been assumed that the extrapolation to low radio frequencies is valid, and the solid angle of the synthesized PSF (with angular FWHM $\theta_\textrm{s}$) is $\Omega = \pi(\theta_\textrm{s}/2)^2$. For the nominal value of $\theta_\textrm{s}=10\arcsec$ chosen in this paper, $\sigma_\textrm{c}(\nu_z,\theta_\textrm{s})\approx 19.9\,\mu$Jy, $\approx 22.2\,\mu$Jy, and $\approx 24.3\,\mu$Jy, respectively in the spectral subbands corresponding to $z=8$, 9.5, and 11. The {\it rms} from sidelobe confusion in each voxel of the image cube is \citep{bow09}:
\begin{align}\label{eqn:sidelobe-confusion-rms}
    \sigma_\textrm{s} &\approx \sigma_\textrm{c} B_\textrm{rms}\left(\frac{\Omega_\textrm{p}}{\Omega}\right)^\frac{1}{2} = \sigma_\textrm{c}B_\textrm{rms}\left(\frac{\theta_\textrm{p}}{\theta_\textrm{s}}\right),
\end{align}
where $B_\textrm{rms}$ is the {\it rms} of the synthesized PSF relative to the peak in any single spectral channel (without bandwidth synthesis), and $\Omega_\textrm{p}=\pi(\theta_\textrm{p}/2)^2$ is the solid angle of the antenna primary beam. 

The sidelobes may contain correlated spectral structure in general. However, a simplifying assumption is made here to obtain an approximate estimate of the power spectrum of the sidelobes from residuals, namely, the sidelobes are uncorrelated along the spectral axis and thus exhibit a ``white'' power spectrum (similar to thermal noise) but restricted to $k_\parallel\lesssim k_{\parallel,\textrm{w}}$.   

Similar to Equation~(\ref{eqn:noise-dspec-rms}), the {\it rms} in each mode of the {\it Fourier} transform of the sidelobes from the residuals across the field of view can be estimated as:
\begin{align}
    \widetilde{\sigma}_\textrm{s}(\hat{\boldsymbol{s}}) &\approx \sigma_\textrm{s}\left(\frac{\Delta r_z}{\delta r_z}\right)^{1/2}\delta r_z,
\end{align}
in units of Jy~Hz, and the corresponding {\it rms} in the 1D power spectrum at $k_\parallel\lesssim k_{\parallel,\textrm{w}}$ as:
\begin{align}
    P_\textrm{s}(\hat{\boldsymbol{s}}) &\approx \frac{\left[\widetilde{\sigma}_\textrm{s}(\hat{\boldsymbol{s}})\right]^2}{\Delta r_z}\left(\frac{\lambda_z^2}{2 k_\textrm{B}\Omega}\right)^2 \approx \sigma_\textrm{s}^2 \,\delta r_z \left(\frac{\lambda_z^2}{2 k_\textrm{B}\Omega}\right)^2,
\end{align}
which is in units of K$^2\,h^{-1}\,$cMpc. Using Equation(\ref{eqn:sidelobe-confusion-rms}) and $P_\textrm{s}\equiv\langle P^2_\textrm{s}(\hat{\boldsymbol{s}})\rangle^{1/2}=P_\textrm{s}(\hat{\boldsymbol{s}})/\sqrt{N_\gamma}$, 
\begin{equation}
    P_\textrm{s} \approx 
    \begin{cases}
    \sigma_\textrm{c}^2 B_\textrm{rms}^2\,\frac{\Omega_\textrm{p}}{\Omega} \left(\frac{\lambda_z^2}{2 k_\textrm{B}\Omega}\right)^2 \frac{\delta r_z}{\sqrt{N_\gamma}}, & k_\parallel\lesssim k_{\parallel,\textrm{w}}\\
    0,              & \text{otherwise}
    \end{cases}
\end{equation}
Similar to the previous sections, by requiring that $\langle P(\hat{\boldsymbol{s}},k_\parallel)\rangle / P_\textrm{s} \ge \rho$ at $k_\parallel\lesssim k_{\parallel,\textrm{w}}$, the requirement on $B_\textrm{rms}$ can be inferred as: 
\begin{align}\label{eqn:B_rms}
    B_\textrm{rms} &\lesssim \left[\frac{\sqrt{N_\gamma}\,\left\langle P(\hat{\boldsymbol{s}},k_\parallel)\right\rangle}{\rho\,\,\sigma_\textrm{c}^2 \,\delta r_z\left(\frac{\lambda_z^2}{2 k_\textrm{B}\Omega}\right)^2}\frac{\Omega}{\Omega_\textrm{p}}\right]^{\frac{1}{2}}, \,\, k_\parallel\lesssim k_{\parallel,\textrm{w}}
\end{align}

\begin{figure*}
\centering
% \subfloat[][EoS model 1 (\tt{BRIGHT GALAXIES})\label{fig:B_rms_bright_gal}]{\includegraphics[width=\textwidth]{figures/Brms_required_Bright_galaxies_fiducial_1024.pdf}} \\
% \subfloat[][EoS model 2 (\tt{FAINT GALAXIES})\label{fig:B_rms_faint_gal}]{\includegraphics[width=\textwidth]{figures/Brms_required_Faint_galaxies_fiducial_1024.pdf}}
\subfloat[][EoS model 1 (\tt{BRIGHT GALAXIES})\label{fig:B_rms_bright_gal}]{\includegraphics[width=\textwidth]{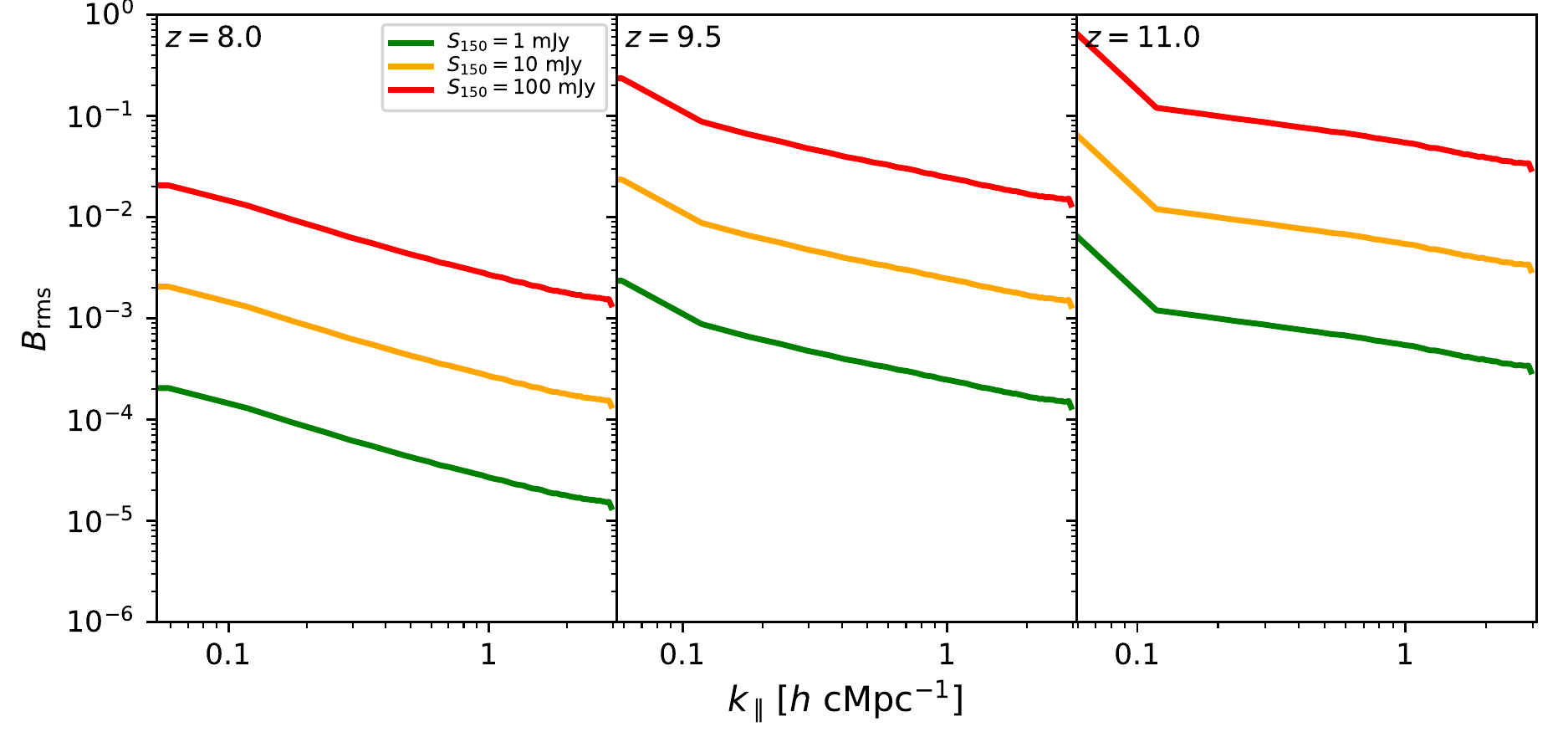}} \\
\subfloat[][EoS model 2 (\tt{FAINT GALAXIES})\label{fig:B_rms_faint_gal}]{\includegraphics[width=\textwidth]{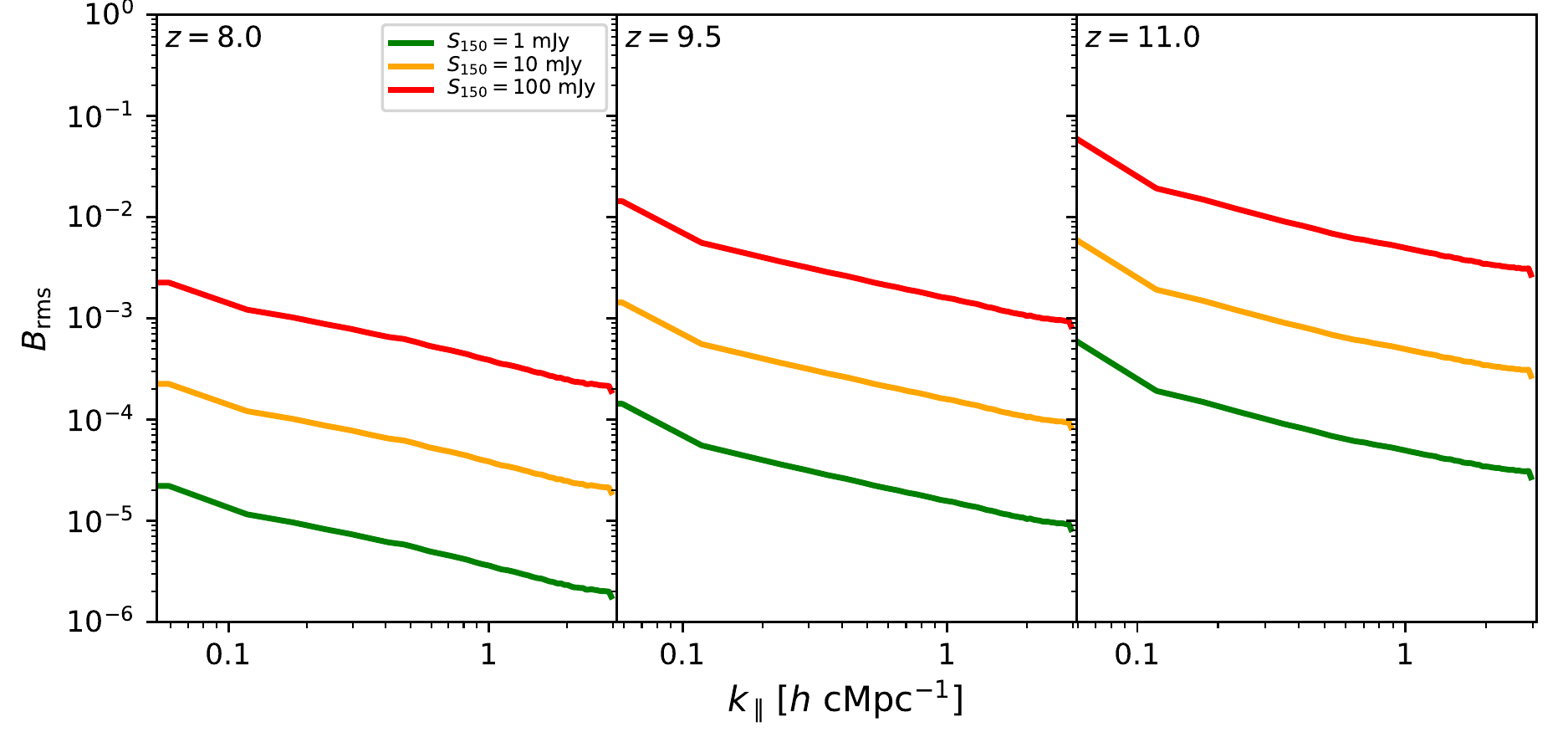}}
\caption{The upper limit on the {\it rms} of the sidelobes per spectral channel in the synthesized PSF, $B_\textrm{rms}$, from Equation~(\ref{eqn:B_rms}), required to control the contamination caused by the chromaticity of the sidelobes emanating from confusing sources in the field of view in order to detect the 1D power spectrum with $\textrm{S/N} \ge \rho$ on every $k_\parallel$-mode at various redshifts and against compact background radiation sources of various strengths using $N_\gamma=100$, field of view $\theta_\textrm{p}=5\arcdeg$, and nominal values for other parameters listed in Table~\ref{tab:nominal-values}. These upper limits apply up to $k_\parallel\lesssim k_{\parallel,\textrm{w}}$, where $k_{\parallel,\textrm{w}}$ denotes the boundary of the {\it foreground wedge} and $k_{\parallel,\textrm{w}}\approx 3.1\,h\,$cMpc$^{-1}$, $\approx 3.4\,h\,$cMpc$^{-1}$, and $\approx 3.6\,h\,$cMpc$^{-1}$ at $z=8$, 9.5, and 11, respectively. At $k_\parallel \gtrsim k_{\parallel,\textrm{w}}$, the sidelobe contamination is expected to be drastically lower and hence, these upper limits on $B_\textrm{rms}$ do not apply. Although the angular resolution has been assumed to be constant at 10\arcsec across the three subbands, the inherent source confusion {\it rms} increases at lower frequencies due to the spectral index. However, the increase in signal power at lower frequencies is even larger and therefore, the upper limit on $B_\textrm{rms}$ becomes less strict at lower frequencies. It has been assumed that the sidelobe structures are spectrally uncorrelated and therefore exhibit flat 1D power up to $k_\parallel\lesssim k_{\parallel,\textrm{w}}$. However, in practice, there will be a spatial correlation that will drop with increasing $k_\parallel$ which will flatten and relax the upper limit requirement on $B_\textrm{rms}$ at higher $k_\parallel$-modes. \label{fig:B_rms}}
\end{figure*}

Figure~\ref{fig:B_rms_bright_gal} and Figure~\ref{fig:B_rms_faint_gal} show the upper limits on the sidelobe {\it rms} required per spectral channel as a fraction relative to the peak for the EoS models 1 and 2 respectively in the different redshift subbands for various values of the compact background source flux strength, $S_{150}$, and nominal values of various parameters listed in Table~\ref{tab:nominal-values}. In other words, the synthesized aperture must be of sufficiently high quality to keep the sidelobes of the synthesized beam below the curves shown in any given spectral channel in the subband in order to keep the sidelobe contamination in the power spectrum contained and achieve a detection above the specified significance threshold, $\rho$. 

Key factors that determine the strength of the sidelobe contamination in the power spectrum are the {\it rms} of the classical radio source confusion, $\sigma_\textrm{c}$, the solid angle of the primary beam, $\Omega_\textrm{p}$, and the sidelobe levels, $B_\textrm{rms}$. $\Omega_\textrm{p}$ determines the number of pixels that contain such confusing sources. Because  $P_\textrm{S}\sim \sigma_\textrm{c}^2\Omega_\textrm{p}$, the ideal criteria for mitigating sidelobe contamination, besides extending the duration of aperture synthesis to achieve a filled aperture, are obtained by an instrument that tends to have a higher angular resolution (smaller $\Omega\sim\theta_\textrm{s}^2$) and a smaller field of view ($\Omega_\textrm{p}\sim\theta_\textrm{p}^2$). 

The significance of the effects of sidelobe contamination is seldom discussed in 21~cm absorption studies during the {\it cosmic dawn} and the EoR because only bright background sources of radiation are typically considered in a direct detection approach. However, most low-frequency instruments typically cover wide fields of view, and this work demonstrates that sidelobe contamination can be significant and should be considered in detail in both the direct detection and power spectrum approaches in wide-field measurements, even for a bright background as seen in the $S_{150}=100$~mJy case. 

Consistent with previous discussions, the overall trend is that the EoS model 2 places a more severe constraint on the quality of the synthesized aperture. The upper limit on $B_\textrm{rms}$ appears less severe at lower-frequency subbands despite the expectation that $\sigma_\textrm{c}$ increases at lower frequencies from Equation~(\ref{eqn:source-confusion-rms}) due to the assumed radio spectral index (note that the angular resolution has been assumed to remain the same across all subbands at 10\arcsec). Although the sidelobe confusion {\it rms} does increase at lower frequencies, the 1D power increases by an even larger factor at higher redshifts (see Figure~\ref{fig:pspec_redshifts}). Therefore, the upper limit requirement on $B_\textrm{rms}$ becomes less severe. It must be emphasized that the requirement on $B_\textrm{rms}$ presented here is pessimistic and depicts a conservative scenario. In practice, unlike the assumption that the sidelobes are spectrally uncorrelated, there is evidence of a nonzero spectral correlation and the power from the sidelobe contamination tends to decrease with increasing $k_\parallel$ \citep[for example, see][]{thy15a}. This will in turn lead to a flattening of the curves in Figure~\ref{fig:B_rms}, implying that the upper limit requirement on $B_\textrm{rms}$ will be less severe and remain closer to the values seen at the lower $k_\parallel$-modes than that presented here. 

\section{Summary}\label{sec:summary}

The nature of radiative transfer presents unique prospects to detect the IGM structures during the {\it cosmic dawn} and the EoR via absorption of the redshifted 21~cm spectral line of H~{\sc i} along directions that contain a compact source of background radiation such as a quasar or AGN, a star-forming radio galaxy, a radio afterglow from a GRB, etc. Previous studies have either limited such prospects to identifying bright background sources owing to sensitivity limitations of the observing instrument for a direct detection in the spectra, explored the change of variance in absorbed regions and elsewhere, or estimated the net statistical effect of such absorption on the full 3D power spectrum. This paper takes a related but unique approach by considering the 1D power spectrum in the direction of such compact background sources of radiation including fainter sources. This approach has the advantage of gaining sensitivity from a power spectrum approach (relative to direct detection) while restricting to only those sightlines where the signals have been boosted in absorption on account of the compact background radiation.

The first half of this paper sets up the theoretical formalism and uses two 21cmFAST models ({\tt BRIGHT GALAXIES} and the {\tt FAINT GALAXIES} models) that span a wide range of astrophysical parameters and yet are plausible based on the data observed to date. Using a simple continuum model for the compact background radiation source (with angular extent assumed to be $\simeq 10\arcsec$ and the flux density observed at 150~MHz parameterized as $S_{150}=1$~mJy, 10~mJy, and 100~mJy), the absorption features even against a relatively faint background source ($S_{150}=1$~mJy) are demonstrated to be much stronger than those with only CMB as the background, resulting in a stronger 1D power spectrum signal. The {\tt BRIGHT GALAXIES} model exhibits stronger power relative to the {\tt FAINT GALAXIES} model. The 1D power increases with redshift (due to increasing optical depth resulting from decreasing spin temperature, and increasing strength of the continuum background radiation toward lower frequencies) in both models thereby raising the prospects of detection at higher redshifts. Hints of studying non-Gaussian features using higher-order moments are also noted in the probability distribution of the flux densities in the voxels in the spectra along sightlines to such compact background objects. 

The second half of the paper addresses the detection prospects as well as requirements on modern low-frequency telescopes such as the LOFAR, EGMRT, SKA1-low, and SKA2 by using generic instrument parameters such as array sensitivity (parameterized by $N_\textrm{a}A_\textrm{e}/T_\textrm{sys}$) and synthesized PSF quality (parameterized by {\it rms} level of sidelobes in the synthesized PSF, $B_\textrm{rms}$) and observing parameters such as number of target background objects beyond a given redshift, $N_\gamma$, and integration time per target, $\delta t$, etc. In general, the 1D power spectrum can significantly improve the prospects of detecting the IGM structures with much reduced cosmic variance and even new spatial scales (particularly smaller scales) that are less accessible with a direct detection approach. 

By requiring that the 1D power spectrum is detectable with a minimum signal-to-noise ratio on all scales, the requirements on $N_\gamma$, $N_\textrm{a}A_\textrm{e}/T_\textrm{sys}$, and $B_\textrm{rms}$ are deduced for both the EoS models at various redshifts for varying background source strengths. While nominal values for parameters that can be generically applied to modern low-frequency radio telescopes were used to deduce $N_\gamma$, $N_\textrm{a}A_\textrm{e}/T_\textrm{sys}$, and $B_\textrm{rms}$, detailed expressions are provided to adjust and scale the requirements to specific instrument and observation details. 

Although the presence of a significant radio population of high-redshift sources is not confirmed by observations yet potentially due to selection biases, the nominal values of $N_\gamma\gtrsim 100$ at different redshifts in this paper appear plausible based on current models. Based on the known and anticipated values of $N_\textrm{a}A_\textrm{e}/T_\textrm{sys}$, all instruments considered here (LOFAR, EGMRT, SKA1-low, and SKA2) are capable of detecting the 1D power spectrum on specific scales, redshifts, and selected values of $S_{150}$ but the detection prospects improve significantly with the SKA1-low ($N_\textrm{a}A_\textrm{e}/T_\textrm{sys}=800\,\textrm{m}^2\,\textrm{K}^{-1}$) and eventually even further with the SKA2 ($N_\textrm{a}A_\textrm{e}/T_\textrm{sys}=4000\,\textrm{m}^2\,\textrm{K}^{-1}$). The effect of contamination from the chromaticity (spectral structure) of PSF sidelobes arising from confusing foreground objects filling the field of view is investigated and found to be important, especially in wide-field measurements. Therefore, achieving a filled synthesized aperture to yield a high-quality synthesized PSF is an important requirement for each of the direct detection, the 1D power spectrum, and the full 3D power spectrum approaches. In general, the observational and instrument requirements tend to become less severe and the detection more likely  at lower frequencies (higher redshifts) due to the inherent increase in the strength of the 1D power spectrum with increasing redshift. 

The line-of-sight 1D power spectrum approach of detecting absorption by neutral IGM structures against a compact background radiation source during the cosmic reionization process not only reduces uncertainties arising from cosmic variance, but also improves sensitivity because of the boosting of absorption signatures by the presence of a potentially large number of compact background radiation sources, especially fainter objects ($S_{150}\lesssim 10$~mJy). The 1D power spectrum along specific narrow sightlines does not suffer from a dilution of the signal power due to cancellations of the signal between emitting and absorbing regions especially on the larger scales, which happens in a 3D power spectrum approach where the CMB is the predominant background. Because wide-field measurements are not a necessity in this approach, it can potentially avoid some of the challenges associated with imaging and analysis of wide-field measurements that are typical of tomography and 3D power spectrum approaches. 

Based on the nominally chosen parameters in a $\lesssim 1000$~hr observing campaign, modern low-frequency interferometer arrays are capable of detecting the 1D power spectrum along the line of sight with high significance to reveal rich information about the H~{\sc i} structures and the astrophysics in the IGM during these cosmic epochs. It presents an independent, complementary, and viable alternative, especially for characterizing the structures on the smallest scales as well as discriminating between \textit{cosmic reionization} models, whereas a direct detection will require enormous sensitivity and observing time or very bright background sources, which are evidently rare. Nevertheless, a direct detection approach yields information on localization along the sightline and other details that are inaccessible in a power spectrum. Therefore, complementary follow-up for direct detection of 21~cm absorption features in the spectrum against any bright background object has significant value.

\acknowledgments

N.T. acknowledges valuable inputs from Nissim Kanekar, Judd Bowman, Aaron Ewall-Wice, Chris Carilli, Mario Santos, and Andrei Mesinger. The National Radio Astronomy Observatory is a facility of the National Science Foundation operated under cooperative agreement by Associated Universities, Inc.

%% Following the acknowledgments section, use the following syntax and the
%% \facility{} or \facilities{} macros to list the keywords of facilities used 
%% in the research for the paper.  Each keyword is check against the master 
%% list during copy editing.  Individual instruments can be provided in 
%% parentheses, after the keyword, but they are not verified.

% \vspace{5mm}
% \facilities{HST(STIS), Swift(XRT and UVOT), AAVSO, CTIO:1.3m,
% CTIO:1.5m,CXO}

\software{21cmFAST \citep{mes11}, Precision Radio Interferometry Simulator \citep[PRISim\footnote{PRISim is publicly available for use under the MIT license at \url{https://github.com/nithyanandan/PRISim}};][]{PRISim_software}, AstroUtils\footnote{AstroUtils is publicly available for use under the MIT license at \url{https://github.com/nithyanandan/AstroUtils}} \citep{AstroUtils_software}, Astropy \citep{astropy:2013,astropy:2018}, NumPy \citep{numpy:2006,numpy:2011}, SciPy \citep{scipy:2020}. }

\appendix

\section{Radio $K$-correction revisited}\label{sec:K-correction}

The relationship between the specific brightness in the observed frame ($z=0$) at observed frequency $\nu_\textrm{o}$ and an arbitrary moving frame at redshift $z$ and some arbitrary frequency $\nu$, is examined assuming the spectrum is described by a power-law spectral index, $\alpha$. The specific brightness, $B_\nu(\nu,z)$, in the two frames is related by \citep{con18}: 
\begin{align}
    B_\nu\left(\frac{\nu}{1+z},z=0\right) \, \frac{\mathrm{d}\nu}{1+z} &=  \frac{B_\nu(\nu,z)}{(1+z)^4}\,\mathrm{d}\nu,
\end{align}
where the term $(1+z)^{-4}$ is known as the {\it cosmological dimming} factor and arises because of factors $(1+z)^2$ each from dependence on the square of the angular diameter and the luminosity distance. Therefore:
\begin{align}
    B_\nu(\nu,z) &= (1+z)^3\,B_\nu\left(\frac{\nu}{1+z},z=0\right). 
\end{align}
In the observed frame, the specific brightness between two frequencies, $\nu/(1+z)$ and $\nu_\textrm{o}$ is related by: 
\begin{align}
    B_\nu\left(\frac{\nu}{1+z},z=0\right) &= \frac{B_\nu(\nu_\textrm{o},z=0)}{(1+z)^\alpha}\left(\frac{\nu}{\nu_\textrm{o}}\right)^\alpha.
\end{align}
Hence, 
\begin{align}\label{eqn:intensity-K-correction}
    B_\nu(\nu,z) &= B_\nu(\nu_\textrm{o},z=0)(1+z)^{3-\alpha}\left(\frac{\nu}{\nu_\textrm{o}}\right)^\alpha. 
\end{align}
Here, both $\nu$ and $\nu_\textrm{o}$ are in general arbitrary and unrelated to each other. This equation represents the full relationship of the specific brightness in the two frames including the K-correction. When $\nu=(1+z)\nu_\textrm{o}$, Equation~(\ref{eqn:intensity-K-correction}) reduces to the more familiar form:
\begin{align}\label{eqn:intensity-redshift}
    B_\nu(\nu_\textrm{o},z=0) &= \frac{B_\nu\left(\nu_\textrm{o}[1+z],z\right)}{(1+z)^3}. 
\end{align}
Even though $\alpha$ does not appear explicitly, it may be implicitly present in $B_\nu(\nu_\textrm{o}[1+z],z)$ and $B_\nu(\nu_\textrm{o},z=0)$ depending on the nature of the radiation. 

A similar relationship between the brightness temperatures in the two frames can be established using the Rayleigh-Jeans approximation, 
\begin{align}
    T(\nu,z) &\approx \frac{c^2}{2\,k_\textrm{B}\,\nu^2} B_\nu(\nu,z) \nonumber\\
    &\approx B_\nu(\nu_\textrm{o},z=0) \left(\frac{\nu}{\nu_\textrm{o}}\right)^\alpha \frac{c^2(1+z)^{3-\alpha}}{2\,k_\textrm{B}\,\nu^2},
\end{align}
where $c$ is the speed of light. Therefore,
\begin{align}\label{eqn:temperature-K-correction}
    T(\nu_\textrm{o},z=0) &\approx \frac{T(\nu,z)}{(1+z)^{3-\alpha}}\left(\frac{\nu}{\nu_\textrm{o}}\right)^{2-\alpha}.  
\end{align}
This is the general expression that relates the brightness temperatures in two different frames and arbitrary frequencies, including the K-correction. When $\nu=(1+z)\nu_\textrm{o}$, Equation~(\ref{eqn:temperature-K-correction}) reduces to:
\begin{align}\label{eqn:temperature-redshift}
    T(\nu_\textrm{o}[1+z],z) &= (1+z)\,T(\nu_\textrm{o},z=0)
\end{align}
Again, even though $\alpha$ does not appear explicitly, depending on the type of radiation, it may be implicitly present in $T(\nu_\textrm{o}[1+z],z)$ and $T(\nu_\textrm{o},z=0)$. 

\section{Power Spectrum of Optical Depth along the Line of Sight}\label{sec:optical-depth-PS}

Realistically, the compact background radiation sources at high redshifts will span a range of luminosities based on a radio luminosity function and thus they will be observed over a range of flux densities. Thus, observations toward each of the compact background radiation sources will appear with different signal-to-noise ratios depending on the strengths of the background objects as illustrated by Figure~\ref{fig:dS_obs_redshifts}. Combining their individual 1D power spectra by simple averaging will have the modulating effects of the background source strength (responsible for the boosting of the strength of the absorption signatures) imprinted in the results and thus make the interpretation of the results complicated. 

One simple approach to marginalize these effects to first order is to estimate the spectrum of the optical depth in Equation~(\ref{eqn:dS_obs_bright_bg}) which has used the simplifying assumptions that smooth continuum spectra have been perfectly removed, $\tau_\nu(\hat{\boldsymbol{s}},\nu)\ll 1$, and $S_\gamma(\hat{\boldsymbol{s}},\nu,z=0) \approx S_\textrm{obs}^\textrm{rad}(\hat{\boldsymbol{s}},\nu,z=0) \gg S_\textrm{s}(\hat{\boldsymbol{s}},\nu,z=0)$ (which is mostly valid as seen from Figure~\ref{fig:fluxdensity_comparison}). Further, ignoring the sky-averaged monopole component, $\langle\delta S(\hat{\boldsymbol{s}},\nu)\rangle$ which is typically absent in an interferometer measurement, the optical depth can be estimated as:
\begin{align}\label{eqn:tau-estimate}
    \tau_\nu^\prime(\hat{\boldsymbol{s}},S_{150},\nu) &\approx -\frac{\Delta S_\textrm{obs}(\hat{\boldsymbol{s}},\nu) }{S_\textrm{obs}^\textrm{rad}(\hat{\boldsymbol{s}},\nu,z=0)}.
\end{align}
It is seen that $\tau_\nu^\prime(\hat{\boldsymbol{s}},S_{150},\nu)\equiv \tau_z^\prime(\hat{\boldsymbol{s}},S_{150},z)$ has normalized the effect of the background radiation, $S_\textrm{obs}^\textrm{rad}(\hat{\boldsymbol{s}},\nu,z=0)$, to first order. Correspondingly, the {\it rms} error in this estimate of optical depth due to thermal noise will be:
\begin{align}\label{eqn:tau-rms-noise}
    \delta\tau_\nu^\textrm{N}(\hat{\boldsymbol{s}},S_{150},\nu) &\approx \frac{\delta S^\textrm{N} }{S_\textrm{obs}^\textrm{rad}(\hat{\boldsymbol{s}},\nu,z=0)},
\end{align}
and that due to synthesized PSF sidelobes will be:
\begin{align}\label{eqn:tau-rms-sidelobes}
    \delta\tau_\nu^\textrm{s}(\hat{\boldsymbol{s}},S_{150},\nu) &\approx \frac{\sigma_\textrm{s} }{S_\textrm{obs}^\textrm{rad}(\hat{\boldsymbol{s}},\nu,z=0)}.
\end{align}

The 1D power spectrum of $\tau_\nu(\hat{\boldsymbol{s}},\nu)$ along the sightline similar to the study of opacity statistics \citep{desh00a,desh00b,lazarian08} could yield direct statistical constraints on the crucial optical depth parameter on different scales at these cosmic epochs. Let the 1D power spectrum of the estimated $\tau_\nu^\prime(\hat{\boldsymbol{s}},S_{150},\nu)$ and the true $\tau_\nu(\hat{\boldsymbol{s}},\nu)$ be denoted by $P_\tau^\prime(\hat{\boldsymbol{s}},S_{150},k_\parallel)$ and $P_\tau(\hat{\boldsymbol{s}},k_\parallel)$, respectively. Analogous to the derivation in \S\ref{sec:pspec}:
\begin{align}\label{eqn:comoving-PS-tau}
    P_\tau^\prime(\hat{\boldsymbol{s}},S_{150},k_\parallel) &\approx \frac{\left|\Delta\widetilde{S}^\prime(\hat{\boldsymbol{s}},k_\parallel)\right|^2}{\left[S_{150}\left(\frac{\nu_z}{\nu_{150}}\right)^\alpha\right]^2} \left(\frac{1}{\Delta r_z}\right) \nonumber\\ 
    &\approx \frac{P(\hat{\boldsymbol{s}},k_\parallel)}{\left[S_{150}\left(\frac{\nu_z}{\nu_{150}}\right)^\alpha\left(\frac{\lambda_z^2}{2k_\textrm{B}\Omega}\right)\right]^2},
\end{align}
with units of $h^{-1}\,$cMpc. The approximation results from assuming that the variation of the continuum background radiation within the spectral subband is ignored and assumed to have a constant value of  $S_{150}(\nu_z/\nu_{150})^\alpha$. The reference $P_\tau(\hat{\boldsymbol{s}},k_\parallel)$ can be derived as the power spectrum of the true optical depth, $\tau_\nu(\hat{\boldsymbol{s}},\nu)$, available from the 21cmFAST simulations. 

Assuming $\tau_\nu(\hat{\boldsymbol{s}},\nu)$, the fluctuations corresponding to $\delta\tau_\nu^\textrm{N}(\hat{\boldsymbol{s}},S_{150},\nu)$ and $\delta\tau_\nu^\textrm{s}(\hat{\boldsymbol{s}},S_{150},\nu)$ are uncorrelated with each other, $P_\tau^\prime(\hat{\boldsymbol{s}},S_{150},k_\parallel)$ can be expressed as: \begin{align}\label{eqn:comoving-PS-tau-net}
    P_\tau^\prime(\hat{\boldsymbol{s}},S_{150},k_\parallel) &= P_\tau(\hat{\boldsymbol{s}},k_\parallel) + \delta P_\tau(\hat{\boldsymbol{s}},S_{150},k_\parallel),
\end{align}
where $\delta P_\tau(\hat{\boldsymbol{s}},S_{150},k_\parallel)$ denotes the fluctuations in the estimated 1D power spectrum of the optical depth. The variance of the uncertainty in the 1D power spectrum from these fluctuations is:
\begin{align}\label{eqn:comoving-PS-tau-error}
    \left\langle\left[\delta P_\tau(\hat{\boldsymbol{s}},S_{150},k_\parallel)\right]^2\right\rangle &= \left\langle\left[P_\tau^\textrm{N}(\hat{\boldsymbol{s}},S_{150},k_\parallel)\right]^2\right\rangle \nonumber\\
    & + \left\langle\left[P_\tau^\textrm{s}(\hat{\boldsymbol{s}},S_{150},k_\parallel)\right]^2\right\rangle \nonumber\\
    & + \left\langle\left[P_\tau^\textrm{e}(\hat{\boldsymbol{s}},S_{150},k_\parallel)\right]^2\right\rangle,
\end{align}
where $P_\tau^\textrm{N}(\hat{\boldsymbol{s}},S_{150},k_\parallel)$ and $P_\tau^\textrm{s}(\hat{\boldsymbol{s}},S_{150},k_\parallel)$ are the {\it rms} fluctuations in the 1D power spectra corresponding to the fluctuations in the line-of-sight voxels caused by thermal noise and sidelobes with {\it rms} $\delta\tau_\nu^\textrm{N}(\hat{\boldsymbol{s}},S_{150},\nu)$ and $\delta\tau_\nu^\textrm{s}(\hat{\boldsymbol{s}},S_{150},\nu)$, respectively. $P_\tau^\textrm{e}(\hat{\boldsymbol{s}},S_{150},k_\parallel)$ denotes the {\it rms} of the error that arises when the assumption made above is invalid, namely, $S_\gamma(\hat{\boldsymbol{s}},\nu,z=0) \not\gg S_\textrm{s}(\hat{\boldsymbol{s}},\nu,z=0)$ or $\tau_\nu(\hat{\boldsymbol{s}},\nu) \not\ll 1$. The angular brackets denote marginalization over multiple independent realizations covering $\hat{\boldsymbol{s}}$ (similar to the discussion in \S\ref{sec:pspec} and \S\ref{sec:pspec-noise}) for fixed values of $k_\parallel$ and $S_{150}$, and thus denote the true variance of the underlying distribution.  

The S/N of the estimated optical depth power spectrum is:
\begin{align}
    \rho_\tau(\hat{\boldsymbol{s}},S_{150},k_\parallel) &= \frac{P_\tau^\prime(\hat{\boldsymbol{s}},S_{150},k_\parallel)}{\left\langle\left[\delta P_\tau(\hat{\boldsymbol{s}},S_{150},k_\parallel)\right]^2\right\rangle^{1/2}}.
\end{align}
In observations, only a discrete sampling rather than a complete two-dimensional distribution of the $\hat{\boldsymbol{s}}$--$S_{150}$ parameter space may be available. In order to marginalize over the distribution of discrete combinations of the observed pairs of $S_{150}$ and $\hat{\boldsymbol{s}}$, a simple scheme towards improving the resulting S/N would be to perform a S/N-weighted average over the pairs:
\begin{align}
    \widehat{P}_\tau(k_\parallel) &= \frac{\sum\limits_{(\hat{\boldsymbol{s}},S_{150})}\,\rho_\tau^2(\hat{\boldsymbol{s}},S_{150},k_\parallel)\,P_\tau^\prime(\hat{\boldsymbol{s}},S_{150},k_\parallel)}{\sum\limits_{(\hat{\boldsymbol{s}},S_{150})}\,\rho_\tau^2(\hat{\boldsymbol{s}},S_{150},k_\parallel)},
\end{align}
ignoring the covariance between errors in the power spectrum measured in the $\hat{\boldsymbol{s}}$--$S_{150}$ parameter plane. 

In this formalism, $\widehat{P}_\tau(k_\parallel)$ is a simple estimator of $P_\tau(k_\parallel) \equiv \langle P_\tau(\hat{\boldsymbol{s}},k_\parallel)\rangle$. The estimator is not guaranteed to be either rigorously optimal or without bias. The intent of this simple formalism is only to show a potential pathway to estimating the power spectrum of the optical depth in this statistical approach using redshifted 21~cm absorption by H~{\sc i} in the IGM against a large number of compact background radiation sources of varying strengths at high redshifts. 

% \bibliography{eor}{}
\bibliographystyle{aasjournal}

%% Include this line if you are using the \added, \replaced, \deleted
%% commands to see a summary list of all changes at the end of the article.
%\listofchanges

\end{document}